\documentclass{new_tlp}
\usepackage{mathptmx}

\usepackage{xr} 
\title[Constraint ASP without Grounding] {Constraint
  Answer Set Programming\\ without Grounding
  \thanks{Work partially supported by EIT Digital
    (\url{https://eitdigital.eu}), MINECO project
    \mbox{TIN2015-67522-C3-1-R} (TRACES), Comunidad de Madrid project
    \mbox{S2013/ICE-2731} \emph{N-Greens Software}, NSF~IIS 1718945,
    and NSF~IIS 1423419.}   
}
\author[J. Arias, M. Carro, E. Salazar, K. Marple and G. Gupta]
{JOAQUIN ARIAS, MANUEL CARRO\\
  IMDEA Software Institute \emph{and}
  Universidad Polit\'ecnica de Madrid\\
  \email{joaquin.arias@\{imdea.org,alumnos.upm.es\}, 
    manuel.carro@\{imdea.org,upm.es\}}
  \and ELMER SALAZAR, KYLE MARPLE, and GOPAL GUPTA\\
  University of Texas at Dallas\\ 
  \email{\{ees101020,kmarple1,gupta\}@utdallas.edu}
}

\usepackage{xspace}

\newcommand{\marpleB}{%
  \nocite{marple2017b}%
  (\citeANP{marple2017b} 2017b)\xspace}
\newcommand{\marpleAB}{%
  \nocite{marple2017a, marple2017b}%
  (\citeANP{marple2017a} 2017a, 2017b)\xspace}
\newcommand{\marpleBgupta}{%
  \nocite{marple2017b,gupta2007coinductive}%
  (\citeANP{marple2017b} 2017b, \citeNP{gupta2007coinductive})\xspace}

\usepackage[pdftex,svgnames,usenames]{xcolor}
\usepackage[
    disable,
    textwidth=0.25\textwidth,textsize=scriptsize]{todonotes}
\jdate{March 2003}
\pubyear{2003}
\pagerange{\pageref{firstpage}--\pageref{lastpage}}
\doi{S1471068401001193}

\usepackage{mathptmx}
\usepackage{dsfont}
\usepackage{xcolor}
\usepackage{comment}
\usepackage[utf8]{inputenc}
\usepackage{wrapfig}
\usepackage{adjustbox}

\usepackage{algorithm2e}
\SetAlFnt{\small} 
\SetKwComment{tpl}{}{}
\SetCommentSty{xCommentSty}
\SetNlSty{mynlfont}{}{} 
\LinesNumbered  
\DontPrintSemicolon 
\RestyleAlgo{algoruled} 

\usepackage{tikz}
\usepackage{amssymb}
\usepackage{multicol}
\usepackage{rotating}
\usepackage[justification=centering]{caption}
\usepackage[hidelinks,pdftex]{hyperref}
\usepackage{relsize}

\definecolor{PrologPredicate}{RGB}{0,0,200}
\definecolor{PrologVar}      {RGB}{145,032,039}
\definecolor{PrologComment}  {RGB}{169,082,044}
\definecolor{PrologOther}    {rgb}{0.2,0.2,0.2}
\definecolor{PrologString}   {rgb}{0.2,0.2,0.2}

\usepackage{listings} 
\usepackage{textcomp}
\newcommand{\code}{\lstinline[style=MyInline]}
\lstdefinestyle{tree}
{
  basicstyle = \small\ttfamily\color{PrologPredicate},
  basewidth = 0.5em,
  moredelim = {[s][\color{PrologString}]{ \{}{\} }},
  moredelim = {*[s][{\color{PrologVar}}]{(}{)}},
  literate     =
  {.\\=.}{{\ \char"5C=\ }}3
  {\\=}{{\ \char"5C=\ }}3
  {.<.}{{\ \#<\ }}4
  {.>.}{{\ \#>\ }}4
  {=}{{\ =\ }}3
  {.=.}{{\ \#=\ }}3
  {.=<.}{{\ \#=<\ }}5
  {.>=.}{{\ \#>=\ }}5
}
\usepackage{etoolbox}
\makeatletter
\patchcmd{\lsthk@SelectCharTable}{)}{`}{}{} 
\makeatother 
\lstdefinestyle{MyInline}
{
  basicstyle = \ttfamily\color{PrologOther},
  breaklines = true,
  breakatwhitespace=true,
  literate =
  {,}{}{0\discretionary{,}{}{,}}
  {|}{{\color{PrologOther}$\mid$}}1
  {\\\{}{{\color{PrologOther}\{}}1
  {\\\}}{{\color{PrologOther}\}}}1
  {[}{{\color{PrologOther}\small[}}1
  {]}{{\color{PrologOther}\small]}}1
  {\\$}{{\color{PrologOther}\$}}1
  {.=.}{{\color{PrologOther}\#=}}3
  {.<.}{{\color{PrologOther}\#<}}3
  {.>.}{{\color{PrologOther}\#>}}3
  {.=<.}{{\color{PrologOther}\#=<}}4
  {.>=.}{{\color{PrologOther}\#>=}}4
  {.\\=.}{{\color{PrologOther}{\char"5C}=}}3
  {\\=}{{\color{PrologOther}\char"5C=}}3
  {?-}{{\color{PrologOther}?-\ }}2
  {:-}{{\color{PrologOther}:-}}3
  {),}{{\color{PrologOther})\hspace{-.2em},}}1
}
\lstdefinestyle{MyProlog}
{
  keywords = {},
  upquote = true,
  basicstyle = \small\ttfamily\color{PrologPredicate},
  basewidth = 0.56em,
  moredelim = {**[s][\color{PrologString}]{'}{'}},
  moredelim = {*[s][\color{PrologVar}]{(}{)}},
  moredelim = {*[s][\color{PrologOther}]{:-}{.}},
  moredelim = {*[s][\color{red}]{/*}{*/}},
  commentstyle = \mdseries\color{PrologComment},
  morecomment=[l]\%,
  morecomment=[s]{/*}{*/},
  literate     =
  {|}{{\color{PrologOther}$\mid$}}1
  {[}{{\color{PrologOther}\small[}}1
  {]}{{\color{PrologOther}\small]}}1
  {\\$}{{\color{PrologOther}\$}}1
  {&(}{{\color{PrologOther}(}}1
  {&)}{{\color{PrologOther})}}1
  {&.}{{.}}0
  {.=.}{{\color{PrologOther}\ \#=\ }}4
  {.<.}{{\color{PrologOther}\ \#<\ }}4
  {.>.}{{\color{PrologOther}\ \#>\ }}4
  {.=<.}{{\color{PrologOther}\ \#=<\ }}5
  {.>=.}{{\color{PrologOther}\ \#>=\ }}5
  {.\\=.}{{\color{PrologOther}\char"5C=}}3
  {\\=}{{\color{PrologOther}\char"5C=}}3
  {=}{{\color{PrologOther}=}}2
  {,}{{\color{PrologOther}\footnotesize,}}1
  {;}{{\color{PrologOther}\footnotesize;}}1,
}
\lstdefinestyle{MyASP}
{
  keywords = {},
  upquote = true,
  basicstyle = \small\ttfamily\color{PrologPredicate},
  basewidth = 0.52em,
  moredelim = {*[s][\color{PrologOther}]{:-}{.}},
  moredelim = {*[s][\color{PrologOther}]{\{}{\}}},
  moredelim = {[s][\color{PrologVar}]{(}{)}},
  commentstyle = \color{PrologComment},
  morecomment=[l]\%,
  literate     =
  {..}{..}2
  {,}{{\color{PrologOther}\footnotesize,}}1
}
\usepackage{ntheorem}

{
\theoremindent=.5cm
\theoremheaderfont{\kern-.5cm\normalfont\slshape}
\theorembodyfont{\normalfont}
\theoremseparator{.}
\theoremprework{\setlength\parindent{0cm}}
\newtheorem{example}{Example}
}

\newcommand\lrul[2]{\ensuremath{#1\ \leftarrow \ #2}}
\newcommand\lque[1]{\ensuremath{\leftarrow\ #1}}

\newcommand{\supp}{}

\newcommand{\mcode}[1]{\ensuremath{{\color{PrologOther}\mathtt{#1}}}}
\newcommand{\cV}{\ensuremath{\overline{\mcode{V}}}\xspace}
\newcommand{\V}{\ensuremath{{\mcode{V}}}\xspace}

\newcommand{\redbeflst}{\vspace*{-.5em}}
\newcommand{\redmidbeflst}{}
\newcommand{\redaftlst}{\vspace*{.5em}}
\newcommand{\redbefsec}{}

\newcommand{\redbefwrap}{}
\newcommand{\reditemize}{}
\newcommand{\redmiditemize}{}

\begin{document}

\maketitle


\begin{abstract}
  Extending ASP with constraints (CASP) enhances its expressiveness
  and performance.  This extension is not straightforward as the
  grounding phase, present in most ASP systems, removes variables and
  the links among them, and also causes a combinatorial explosion in
  the size of the program.  
  Several methods to overcome this issue have been devised:
  restricting the constraint domains (e.g., discrete instead of
  dense), or the type (or number) of models that can be returned.  In
  this paper we propose to incorporate constraints into s(ASP), a
  goal-directed, top-down execution model which implements ASP while
  retaining logical variables both during execution and in the answer
  sets.  The resulting model, s(CASP), can constrain variables that,
  as in CLP, are kept during the execution and in the answer sets.
  s(CASP) inherits and generalizes the execution model of s(ASP) and
  is parametric w.r.t.\ the constraint solver.  We describe this novel
  execution model and show through several examples the enhanced
  expressiveness of s(CASP) w.r.t.\ ASP, CLP, and other CASP systems.  We also report improved performance w.r.t.\ other
  very mature, highly optimized ASP systems in some benchmarks. %

  \smallskip\noindent
  This paper is under consideration for publication in Theory and
  Practice of Logic Programming (TPLP).
\end{abstract}

\redbefsec
\section{Introduction}
\label{sec:introduction}

Answer Set Programming (ASP) has emerged as a successful paradigm for
developing intelligent applications.  It uses the stable model
semantics~\cite{gelfond88:stable_models} for 
programs with negation. 
ASP has attracted much attention due to its
expressiveness, ability to incorporate
non-monotonicity, represent knowledge, and model combinatorial 
problems.  On the other hand,
constraints have been used both to enhance
expressiveness and to increase performance  in logic programming.  Therefore, it is
natural to incorporate constraints in ASP systems. This is however not
straightforward as ASP systems usually carry out an initial grounding
phase where variables (and, therefore, the constraints linking them)
disappear. Several approaches have been devised to work around 
this issue.
However, since constraints need to be grounded as well, 
these approaches 
limit the range of admissible constraint domains (e.g., discrete
instead of dense), the places where constraints can appear, and the
type (or number) of models that can be returned. The integration of
constraints with ASP
is not as seamless as in standard constraint logic programming (CLP).

In this work we propose to restore this integration
by incorporating constraints into the
s(ASP)~\marpleB execution model.  s(ASP) is a
goal-directed, top-down, SLD resolution-like procedure which evaluates
programs under the 
ASP semantics without a grounding phase either before or during execution.  
s(ASP) supports predicates and thus retains logical variables both
during execution  
and in the answer sets.  We
have extended s(ASP)'s execution model to make its integration with generic
constraint solvers possible. The resulting execution model and system,
called s(CASP), makes it possible to express constraints on variables and 
extends s(ASP)'s in the same way that CLP extends 
Prolog's execution model. Thus,
s(CASP) inherits and generalizes the execution model of s(ASP) while remaining
parametric w.r.t.\ the constraint solver. Due to its basis in s(ASP), s(CASP)
avoids grounding the program and the concomitant combinatorial explosion.
s(CASP) can also handle answer set programs that manipulate arbitrary 
data structures as well as reals, rationals, etc.
We show, through several examples, its
enhanced expressiveness w.r.t.\ ASP, CLP, and other ASP
systems featuring constraints. We briefly discuss s(CASP)'s efficiency: on
some benchmarks it can outperform
mature, highly optimized ASP systems.

Several approaches have been proposed to mitigate the impact of the
grounding phase in ASP systems.  In the case of large data sets,
\emph{magic set} techniques have been used to improve grounding for
specific queries~\cite{alviano12-magic-sets}.  For programs which use
uninterpreted function symbols, 
techniques such as \emph{external
  sources}~\cite{calimeri07-external-sources} have been proposed.

Despite these approaches, grounding is still an issue when constraints
are used in ASP.  Variable domains induced by constraints
can be unbound and, therefore, infinite (e.g., \code{X.>.0} with
$\mathtt{X} \in \mathds{N}$ or $\mathtt{X} \in \mathds{Q}$).
Even if they are bound, they can contain an infinite number of
elements (e.g. \code{X.>.0 $\land$ X.<.1} in $\mathds{Q}$ or $\mathds{R}$).
These problems have been attacked using different techniques:

\begin{itemize}
\item Translation-based methods~\cite{balduccini17-ezcsp}, which
  convert both ASP and constraints into a theory that is executed
  in an SMT solver-like manner.  Once the
  input program is translated, they benefit from the features and
  performance of the target ASP and CLP solvers. However, the translation may
  result in a large propositional representation or weak propagation
  strength.
\item Extensions of ASP systems with constraint
  propagators~\cite{banbara17-clingcon3,janhunen2017clingo} 
  that generate and propagate new constraints during the search and
  thus continuously check for consistency using external solvers using
  e.g.\ conflict-driven clause learning.
  However, they are restricted to finite domain solvers (hence,
  dense domains cannot be appropriately captured) and
  incrementally generate ground models, lifting the upper bounds for
  some parameters. This, besides being a performance bottleneck, falls
  short of capturing the true nature of variables in constraint
  programming. 
\end{itemize}

Due to the requirement to ground the program, causing a loss of
communication  from elimination of variables, the execution
methods for CASP systems are complex. Explicit hooks sometimes are
needed in the language, e.g., the \texttt{required} builtin of EZCSP~\cite{balduccini17-ezcsp},
so that the ASP solver and the constraint solver can communicate.
Additionally, considerable research has been
conducted on devising top-down execution models for
ASP~\cite{dal2009gasp,baselice09-finitely-recursive,baselice10-decidable-finitary}
that could be extended with constraints.

We have validated the s(CASP) approach with an implementation in Ciao
Prolog which integrates Holzbaur's
CLP($\mathds{Q}$)~\cite{holzbaur-clpqr}, a linear constraint solver
over the rationals.\footnote{Note that while we used CLP($\mathds{Q}$)
  in this paper, CLP($\mathds{R}$) could also have been used.}  The
s(CASP) system has been used to solve a series of problems that
would cause infinite recursion in other top-down systems, but which in
s(CASP) finitely finish, as well as others that require
constraints over dense and/or unbound domains.
Thus, s(CASP) is able to solve problems that cannot be
straightforwardly solved in other systems.

\redbefsec
\section{Background: ASP and s(ASP)}
\label{sec:preliminaries}

\textbf{ASP}~\cite{gelfond88:stable_models,brewka2011answer} is a
logic programming and modelling language. An ASP program $\Pi$ is a
finite set of \emph{rules}. Each rule $r \in \Pi$ is of the form:

\redmidbeflst

\begin{displaymath}
  a \leftarrow b_1 \land \dots \land b_m \land not\ b_{m+1}
\land \dots \land not\ b_n. 
\end{displaymath}

  \medskip

\noindent
where $a$ and $b_1, \dots, b_n$ are atoms and $not$ corresponds to
\emph{default} negation. An atom is an expression of form
$p(t_1,\dots,t_n)$ where $p$ is a predicate symbol of arity $n$ and
$t_i$, are \emph{terms}. An atom is \emph{ground} if no variables
occur in it. The set of all \emph{constants} appearing in $\Pi$ is
denoted by $C_{\Pi}$.  
The head of rule $r$ is $h(r) = \{a\}$\footnote{Disjunctive ASP
  programs (i.e., programs with disjunctions in the heads of rules) can be
  transformed into non-disjunctive ASP programs by using
  \emph{default} negation~\cite{ji2016eliminating}.}  and the
body consists of positive atoms $b^+(r) = \{b_1,\dots,b_m\}$ and
negative atoms $b^-(r)=\{b_{m+1},\dots,b_n\}$. Intuitively, rule $r$
is a justification to \emph{derive} that $a$ is true if all atoms in
$b^+(r)$ have a derivation and no atom in $b^-(r)$ has a
derivation. An interpretation $I$ is a subset of the program's
Herbrand base and it is said to satisfy a rule $r$ if $h(r)$ can be
derived from $I$.  A model of a set of rules is an interpretation that
satisfies each rule in the set.  An answer set of a program $\Pi$ is a
minimal model (in the set-theoretic sense) of the program

\redmidbeflst

\begin{displaymath}
   \Pi^I = \{h(r) \leftarrow b^+(r)\ |\ r\in \Pi, b^-(r) \cap I = \emptyset \} 
\end{displaymath}

  \medskip

\noindent
which is called the \emph{Gelfond-Lifschitz reduct} of $\Pi$ with
respect to $I$~\cite{gelfond91:clasical_negation}. The set of all
answer sets of $\Pi$ is denoted by $AS(\Pi)$. ASP solvers which
compute the answer sets of non-ground programs use the above
semantics by first applying, to each rule $r \in \Pi$, all possible
substitution from the variables in $r$ to elements of $C_{\Pi}$ (this
procedure is called \emph{grounding}). To make this grounding possible, the rules
of the program should be \emph{safe}, i.e., all variables that appear in a
rule have to appear in some positive literal in the body.  The rule is
termed \emph{unsafe} otherwise.

A difference between ASP and Prolog-style (i.e., SLD resolution-based)
languages is the treatment of negated literals.  Negated literals in a
body are treated in ASP using their logical semantics based on
computing stable models.  The \emph{negation as failure} rule of
Prolog (i.e., SLDNF resolution~\cite{Clark78}) makes a negated call
succeed (respectively, fail) iff the non-negated call fails
(respectively, succeeds).  To ensure soundness, SLDNF has to be
restricted to ground calls, as a successful negated goal cannot return
bindings.
However, SLDNF increases the cases of non-termination w.r.t.\ SLD.

\textbf{s(ASP)}~\marpleAB is a
top-down, goal-driven interpreter of ASP programs written in Prolog
(\url{http://sasp-system.sourceforge.net}).  The top-down
evaluation makes the \emph{grounding} phase unnecessary.  The
execution of an s(ASP) program starts with a \emph{query}, and each
answer is the resulting \emph{mgu} of a successful derivation, its
justification, and a (partial) stable model. This partial stable model
is a subset of the ASP stable model~\cite{gelfond88:stable_models}
including only the literals necessary to support the query with its
output bindings.\footnote{Note that the subset property holds only
  when the Gelfond--Lifschitz transformation is applied assuming an
  \emph{extended Herbrand Base} obtained by extending the
   set of constants in the program, $C_{\Pi}$, with an infinite number of
  new elements.}

  \begin{example}[Assuming an extended Herbrand Base] Given the program below:

\redbeflst

    \begin{center}
      \begin{minipage}{.9\linewidth}
\begin{multicols}{2}
\begin{lstlisting}[style=MyProlog]
married(john).
:- not married(X).
\end{lstlisting}
        \end{multicols}
      \end{minipage}
    \end{center}

  \medskip        

\noindent
    most ASP systems are not able to compute its stable  model (not
    even an empty one), because the global constraint is 
    unsafe. On the other hand, s(ASP) is able to compute queries to
    programs with unsafe rules by assuming that the
    unsafe variables take values in an \emph{extended Herbrand Universe}, and not
    just that of the terms which can be constructed from the symbols
    in the program. Therefore, using this alternative semantics
    \code{:-not married(X).}  corresponds to
    $\neg \exists x. \neg married(x) \equiv \forall x.married(x)$ and
    since the program only has evidence of one married individual
    (\code{john}), there is no stable model (i.e., it cannot be
    derived that all possible individuals are married).
    However, if we add the (unsafe) fact \code{married(X)} (i.e.,
    $\forall x.married(x)$) to the program, the resulting stable model
    will be \code{\{married(X)\}} --- every element of the universe is
    married.
  \end{example}

\noindent
s(ASP) has two additional relevant differences w.r.t.\ Prolog: first,
s(ASP) resolves negated atoms $not\ l_i$ against \emph{dual rules} of
the program (Section~\ref{sec:dual-logic-program}), instead of using
negation as failure. This makes it possible for a non-ground negated
call \code{not p(X)} to return the results for which the positive
call \code{p(X)} would fail. 
Second, and very important, the dual program is \textbf{not} interpreted under SLD
semantics: a number of very relevant changes related to how loops are
treated (see later) are introduced.

  \redbefsec
  \subsection{Dual of a Logic Program}
\label{sec:dual-logic-program}

The dual of a predicate \code{p/1} is another predicate that returns
the \code{X} such that \texttt{p(X)} is not true.  It is used to
give a constructive answer to a goal \code{not p(X)}.  The dual of
a logic program is another logic program containing the dual of each
predicate in the \mbox{program~\cite{alferes2004abduction}.}
To synthesize the dual of a logic program $P$ we first obtain 
 Clark's
completion~\cite{Clark78}, which assumes that the
  rules of the program completely capture all possible ways for atomic
  formulas to be true, and then we apply De Morgan's laws:

  \reditemize
  \begin{enumerate}
\item For each literal $p/n$ that appears in the head of a rule, 
  choose a tuple $\vec{x}$ of $n$ distinct, new variables
  $x_1,\ \dots,\ x_n$.
\item For each $i$-th rule of a predicate $p/n$ of the form
  \lrul{p_i(\vec{t_i})}{B_i}, with $i=1,\dots,k$, make a list
  $\vec{y_i}$ of all variables that occur in the body $B_i$ but do not
  occur in the head $p_i(\vec{t_i})$, add $\exists \vec{y_i}$ to the
  body and rename the variables that appear in the head $\vec{t_i}$
  with the tuple $\vec{x}$, obtained in the previous step, resulting
  in a predicate representing
  $\forall\vec{x}\ ( p_i(\vec{x}) \ \leftarrow \ \exists \vec{y_i} \
  B_i)$.  Note that $\vec{x}$ are local, fresh variables.
  This step captures the standard semantics of Horn clauses.

\item With all these rules and using Clark's completion, we
  form the sentences:

    \smallskip

  \begin{tabular}{lll}

    $\forall\vec{x}\ (\ p(\vec{x})$&$\longleftrightarrow$&$p_1(\vec{x})
    \lor \dots \lor p_k(\vec{x})\ )$\\

    $\forall\vec{x}\ (\ p_i(\vec{x})$&$\longleftrightarrow$&$\exists
    \vec{y_i} \ (b_{i.1} \land \dots \land b_{i.m} \land \lnot\ b_{i.m+1}
    \land \dots   \land \lnot\ b_{i.n})\ )$\\

  \end{tabular}

  \medskip

\item Their semantically equivalent duals $\lnot p/n$, $\lnot p_i/n$ are:

  \smallskip

  \begin{tabular}{lll}
    $\forall\vec{x}\ (\ \lnot p(\vec{x})$&$\longleftrightarrow$&$\lnot (
    p_1(\vec{x})
    \lor \dots \lor p_k(\vec{x}))\ )$\\

    $\forall\vec{x}\ (\ \lnot p_i(\vec{x})$&$\longleftrightarrow$&$\lnot \
    \exists \vec{y}_i\ (b_{i.1} \land \dots \land b_{i.m} \land \lnot\
    b_{i.m+1} \land \dots \land \lnot\ b_{i.n})\ )$\\
  \end{tabular}

  \medskip
  
\item Applying De Morgan's laws we obtain:

  \smallskip

  \begin{tabular}{lll}
   $\forall\vec{x} (\ \lnot p(\vec{x})$&$\longleftrightarrow$&$\lnot
    p_1(\vec{x})
    \land \dots \land \lnot p_k(\vec{x})\ )$\\

    $\forall\vec{x}\ ( \lnot p_i(\vec{x})$&$\longleftrightarrow$&$ 
    \forall \vec{y}_i\ (\lnot b_{i.1} \lor \dots \lor \lnot \ b_{i.m}
    \lor \ b_{i.m+1} \lor \dots \lor \ b_{i.n})\ )$
  \end{tabular}

  \medskip

  \noindent which generates a definition for
  $ \lnot p(\vec{x})$ and a separate clause with head $\lnot p_i(\vec{x})$
  for each positive or negative literal $b_{i.j}$ in the disjunction.
  Additionally, a construction to implement the universal quantifier
  introduced in the body of the dual program is necessary
  (Section~\ref{sec:forall2-algorithm}).

 \redaftlst

\end{enumerate}
\reditemize
  
  Definitions for the initially negated literals
  $\lnot b_{i.m+1} \dots \lnot b_{i.n}$ and for each of the \emph{new}
  negated literals $\lnot b_{i.1} \dots \lnot b_{i.m}$ are similarly
  synthesized.  At the end of the chain, unification has to be negated
  to obtain disequality, e.g., $x = y$ is transformed into $x \not= y$
  (Section~\ref{sec:constr-negat}).

\begin{example}
  Given the program below:

  \begin{center}
    \begin{minipage}{1\linewidth}
      \redbeflst
\begin{multicols}{2}
\begin{lstlisting}[style=MyProlog]
p(0).
p(X) :- q(X), not t(X,Y).
q(1).
t(1,2).
\end{lstlisting}
      \end{multicols}
    \end{minipage}
  \end{center}

  \medskip        

  \noindent
  the resulting  dual program is:

  \begin{center}
    \begin{minipage}{1\linewidth}
      \redbeflst
\begin{multicols}{2}
\begin{lstlisting}[style=MyProlog]
not p(X) :- not p1(X), not p2(X).
not p1(X) :- X\=0.
not p2(X) :- 
   forall(Y, not p2_(X,Y)).
not p2_(X,Y) :- not q(X).
not p2_(X,Y) :- q(X), t(X,Y).
not q(X) :- not q1(X).
not q1(X) :- X\=1.
not t(X,Y) :- not t1(X,Y).
not t1(X,Y) :- X\=1.
not t1(X,Y) :- X=1, Y\=2.
\end{lstlisting}
      \end{multicols}
    \end{minipage}
  \end{center}

  \medskip

\end{example}

  \noindent
  For efficiency, the generation of the dual diverges slightly from
  the previous scheme.  The dual of a body $B \equiv l_1 \land \dots$
  is the disjunction of its negated literals
  $\lnot B \equiv \lnot l_1 \lor \dots$, which generates
  independent clauses in the dual program.  To avoid redundant
  answers, every clause for a negated literal
  $\lnot l_i$ includes calls to any positive literal $l_j$ with
  $j < i$.
  E.g., clause 6 from the previous program, %
  \code{not p2(X,Y):- q(X), t(X,Y)}, would only need to be %
  \mbox{\code{not p2(X,Y):- t(X,Y)}}\quad.  However, the literal
  \code{q(X)} is included to avoid exploring solutions already
  provided by clause 5, \mbox{\code{not p2(X,Y):- not q(X)}}\enspace.  The
  same happens with clauses 10 and 11.

  \redbefsec
  \subsection{Constructive Disequality}
\label{sec:constr-negat}

Unlike Prolog's \emph{negation as failure},
disequality in s(ASP),
denoted by ``\verb'\=''' , represents the constructive negation of the
unification and is used to construct answers from negative literals.
Intuitively, \code{X\=a} means that 
\texttt{X} can be any term not unifiable with \texttt{a}.
In the implementation reported in~\marpleB a
variable can only be disequality-constrained against ground terms, and
the disequality of two compound terms may 
require backtracking to check all the cases:
\code{p(1,Y) \= p(X,2)} first succeeds with
\code{X\=1} and then, upon backtracking, with
\code{Y\=2}.

The former restriction reduces the range of valid programs, but this
does not seem to be a problem in practice: 
since positive literals are called before %
negative literals in the dual program,
the number of cases where this situation may occur is further reduced.
Since this is orthogonal to the
implementation framework, it can be improved upon separately.
The second characteristic impacts performance, but can again be
ameliorated with a more involved implementation of disequality which
carries a disjunction of terms.

\redbefsec
\subsection{\texttt{Forall} Algorithm}
\label{sec:forall2-algorithm}

In~\marpleB the universal quantifier is evaluated
by \code{forall(V, Goal)} which checks if \code{Goal} is true for all
the possible values of \code{V}. When \code{forall/2} succeeds,
the evaluation continues with the quantified variable unbound. Multiple
quantified variables are handled by nesting: $\forall v_1, v_2.Goal$ is
executed as %
\code{forall(V1,forall(V2,Goal))}. The underlying idea 
is to verify that
for any solution with \code{V\=a} (for some \code{a}), \code{Goal}
also succeeds with \code{V=a} (Algorithm~\ref{alg:forall}).

\reditemize

\begin{algorithm}[t]

  \emph{forall} receives \code{V}, a variable name, and \code{Goal}, a
  callable goal.

  \code{V} starts unbound
  
  Execute \code{Goal}. %

  \uIf(\tpl*[f]{\emph{Let us check the bindings of \code{V}}}){\code{Goal} succeeded}{%
    \label{item:a}
    
    \lIf(\tpl*[f]{\emph{\code{Goal}'s success is independent of
        \code{V}}}){\code{V} is unbound}{\emph{forall} \textbf{
        succeeds}}

    \lElseIf {\code{V} is bound,}{backtrack to
      step~\ref{item:a} and try other clauses}

    \Else(\tpl*[f]{\emph{\code{V} has been constrained to be
        different from a series of values}}){%

      Re-execute \code{Goal}, successively substituting the
      variable \code{V} with each of these values

      \lIf{ \code{Goal} succeeds for each value}{ \emph{forall}
        \textbf{succeeds}}
      
      \lElse( \tpl*[f]{\emph{There is at least one value for which
          \code{Goal} is not true}}){\emph{forall}
        \textbf{fails}}

    } 

  } 

  \lElse(\tpl*[f]{\emph{There are infinitely many values for which \code{Goal}
      is not true}}){\emph{forall} \textbf{fails}} 

  \caption{\emph{forall}}
  \label{alg:forall}
\end{algorithm}

\begin{example}
  Consider the following program with the dual rule for \code{p/0}:

  \begin{center}
    \hspace{-4em}
    \begin{minipage}{.7\linewidth}
      \redbeflst
\begin{multicols}{2}
\begin{lstlisting}[style=MyProlog]
p :- not q(X).                       
q(X) :- X=a.                       
q(X) :- X\=a.                      
not p :- forall(X, not p1(X)).  
not p1(X) :- q(X).              
\end{lstlisting}
      \end{multicols}
    \end{minipage}
  \end{center}

  \medskip        

  Under the query \code{?- not p}, the interpreter will execute
  \code{forall(X, not p1(X))} with \code{X} unbound. First, %
  \code{not p1(X)} is executed and calls \code{q(X)}, succeeding with
  \code{X=a}.  Then, since \code{X} is bound, the interpreter
  backtracks and succeeds with \code{X\=a} (second clause of
  \code{q/1}).  Now, since \code{X} is constrained to be different
  from \code{a}, the interpreter re-executes \code{not p1(X)} with
  \code{X=a} which succeeds (first clause of \code{q/1}).
  Since  there are no more constrained values to be checked, the
  evaluation of the query finishes with success.
  Note that leaving \code{X} unbound after the success of
  \code{forall(X, p(X))} is consistent with the interpretation that
  the answer set \code{\{p(X)\}} corresponds to $\forall x.p(x)$.
\end{example}

\redbefsec
\subsection{Non-Monotonic Checking Rules}
\label{sec:non-monotonic-rules}

Non-monotonic rules are used by s(ASP) to ensure that partial
stable models are consistent with the global constraints of the program.
Given a consistency rule of the form
$\forall\vec{x} ( p_i(\vec{x}) \leftarrow \exists \vec{y} \ B_i \land
\lnot p_i(\vec{x}) )$, and in order to avoid contradictory rules of
the form $p_i(\vec{a}) \leftarrow \lnot p_i(\vec{a})$, all stable
models must satisfy that at least one literal in $B_i$ is false (i.e.,
$\lnot B_i$) or, for the values $\vec{a}$ where $B_i$ is true,
$p_i(\vec{a})$ can be derived using another rule.  To ensure that
the partial stable model is consistent, the s(ASP) compiler 
generates, for each consistency rule, a rule of the form:

\smallskip
\begin{tabular}{rl}
  $\forall \vec{x} (\ chk_i(\vec{x})\  \longleftrightarrow$
  &
    $ \forall \vec{y}_i (\lnot B_i \lor p(\vec{x}) \ )\ )$ \\
\end{tabular}

\medskip

\noindent
To ensure that each sub-check ($chk_i$) is satisfied, the compiler
introduces into the program the rule
$nmr\_check \leftarrow chk_1 \land \dots \land chk_k$, which is
transparently called after the program query.

\begin{example}
  Given the program below:

  \begin{center}
    \begin{minipage}{.9\linewidth}
      \redbeflst
\begin{multicols}{2}
\begin{lstlisting}[style=MyProlog]
:- not s(1, X).
p(X):- q(X), not p(X).
\end{lstlisting}
      \end{multicols}
    \end{minipage}
  \end{center}
  
  the resulting \emph{NMR} check rules are:

    \redbeflst

  \begin{center}
  \begin{minipage}{.9\linewidth}
\begin{multicols}{2}
\begin{lstlisting}[style=MyProlog]
nmr_check :- 
    chk1, 
    forall(A, chk2(A)).
chk1 :- forall(X,s(1,X)).
chk2(X) :- not q(X).
chk2(X) :- q(X), p(X).
\end{lstlisting}
  \end{multicols}
\end{minipage}
\end{center}

\end{example}

\noindent
\paragraph{\textbf{Infinite Loops}}
Finally, in order to break infinite loops, s(ASP) uses three
techniques to deal with \emph{odd loops over negation}, \emph{even
  loops over negation}, and \emph{positive
  loops}~\marpleBgupta.  Since they are not essential
for this paper, a summary is included in \ref{app:detect-loops}\supp,
for the reader's convenience.

\redbefsec
\section{s(CASP): Design and
 Implementation}
\label{sec:scasp:extend-sasp}

S(CASP) (available together with the benchmarks used in this paper at
\url{https://gitlab.software.imdea.org/joaquin.arias/sCASP}) extends
s(ASP) by computing partial stable models of programs with
constraints.  This extension makes the following contributions:

    \begin{table} 
      \redbefwrap\normalsize
      \setlength{\tabcolsep}{1.75em}
      \begin{tabular}{@{\extracolsep{0pt}}lrr}
        & \textbf{s(CASP)} & \textbf{s(ASP)}  \\ \cline{1-3}
        hanoi(8,T)    & \textbf{1,528} &  13,297  \\
        queens(4,Q)   & \textbf{1,930} &  20,141  \\ 
        One hamicycle & \textbf{493}   &  3,499   \\ 
        Two hamicycle & \textbf{3,605} &  18,026  \\ \cline{1-3}
      \end{tabular}
      \caption{Speed comparison: s(CASP) vs.\ s(ASP) (time in ms).}
      \label{tab:eval-sasp}
    \end{table}

\reditemize
\begin{itemize}
\item

    The interpreter is reimplemented in Ciao
    Prolog~\cite{hermenegildo11:ciao-design-tplp}.  The
    driving design decision of this reimplementation is to let Prolog take care
    of all operations that it can handle natively, instead of
    interpreting them.  Therefore, a large part of the environment for
    the s(CASP) program is carried implicitly in the Prolog
    environment.  Since s(CASP) and Prolog shared many characteristics
    (e.g., the behavior of variables), this results in flexibility of
    implementation (see the interpreter code sketched in
    Figure~\ref{fig:interpreter} and in full in
    \ref{app:scasp-interpreter}\supp) and gives a large performance
    improvement (Table~\ref{tab:eval-sasp}).  Note that all the
    experiments in this paper were performed on a MacOS 10.13 machine
    with an Intel Core i5 at 2GHz. %

  \item A new solver for disequality constraints.

\item The definition and implementation of a generic interface to
  plug-in different constraint solvers.  This required, in addition to
  changes to the interpreter, changes to the compiler which
  generates the dual program.  This interface has been used, in this
  paper, to connect both the disequality constraint solver and the
  CLP($\mathds{Q}$) solver.

\item The design and implementation of \emph{C-forall}
  (Algorithm~\ref{alg:forall2}), a generic algorithm which extends the
  original \emph{forall} algorithm (Algorithm~\ref{alg:forall}) with
  the ability to evaluate goals with variables constrained under
  arbitrary constraint domains.  In addition to being necessary to deal with
  constraints, this extension generalizes and clarifies the design of the
  original one.
\end{itemize}
\reditemize

\redbefsec
\subsection{s(CASP) Programs}
\label{sec:semantics-scasp}

An s(CASP) program is a finite set of rules of the form:

\redmidbeflst

\begin{displaymath}
  a \leftarrow c_a \land b_1 \land \dots \land b_m \land not\ b_{m+1}
\land \dots \land not\ b_n. 
\end{displaymath}

  \medskip

\noindent
where the difference w.r.t.\ an ASP program is $c_a$, a simple constraint or
a conjunction of constraints.  A query to an s(CASP) program is of the
form \lque{c_q \land l_1 \land \dots \land l_n}, where $c_q$ is also a
simple constraint or a 
conjunction of constraints. The semantics of s(CASP) extends that of
s(ASP) following~\cite{survey94}.  During the evaluation of an s(CASP)
program, the interpreter generates constraints whose consistency
w.r.t.\ the current constraint store is checked by the
\emph{constraint solver}.  The existence of variables both during
execution and in the final models is intuitively justified by adopting
an approach similar to that of the
S-semantics~\cite{gabbrielli1991modeling}.

\redbefsec
\subsection{The Interpreter and the Disequality
            Constraint Solver}
\label{sec:interpreter}

The s(CASP) interpreter carries the environment (the call path and the
model) implicitly and delegates to Prolog all operations that Prolog
can do natively, such as handling the bindings due to unification, the
unbinding due to backtracking, and the operations with constraints,
among others.  The clauses of the program, their duals, and the NMR-checks
are created by the compiler by generating rules of the predicate
\code{pr_rule(Head,Body)}, where \code{Head} is an atom and
\code{Body} is the list of literals.  While the s(CASP) interpreter
performs better than s(ASP), little effort has been
invested in optimizing it (see Section~\ref{sec:concl-future-work}),
and there is ample room for improvement.

\begin{figure}
\begin{multicols}{2}
\begin{lstlisting}[style=MyProlog]
??(Query) :-
   solve(Query,[],Mid),
   solve_goal(nmr_check,Mid,Out),
   print_just_model(Out).
solve([], In, ['\$success'|In]).
solve([Goal|Gs], In, Out) :-
   solve_goal(Goal, In, Mid),
   solve(Gs, Mid, Out).
solve_goal(Goal, In, Out) :-
   user_defined(Goal), !,
   pr_rule(Goal, Body),
   solve(Body, [Goal|In], Out).
solve_goal(Goal, In, Out) :-
   call(Goal),
   Out=['\$success',Goal|In].
\end{lstlisting}
  \end{multicols}
  \caption{(Very abridged) Code of the s(CASP) interpreter.}
\label{fig:interpreter}
\redbefsec
\end{figure}

Figure~\ref{fig:interpreter} shows a highly simplified sketch of the
code that implements the interpreter loop in
s(CASP), where:
\begin{itemize}
\item \code{??(+Query)} receives a query and prints the successful
  path derivations.
\item \code{solve(+Goals,+PathIn,-PathOut)} reproduces SLD resolution.
\item \code{solve_goal(+Goal, +PathIn,-PathOut)} evaluates the
  user-defined predicates and hands over to Prolog the execution of
  the builtins using \code{call/1}. The \code{PathOut} argument
  encodes the derivation tree in a list.
\end{itemize}

\noindent
Every
\verb+`$success'+ constant denotes the success of the goals in the
body of a clause and means that one has to go up one level in the
derivation tree.  Several \verb+`$success'+ constants in a row mean,
accordingly, that one has to go up the same number of levels.

In s(CASP), constructive disequality is handled by a disequality
constraint solver, called CLP($\not=$), implemented using attributed
variables that makes disequality handling transparent to the user
code.
The current implementation of CLP($\not=$) does not address the
restrictions described in Section~\ref{sec:constr-negat}; however, as
mentioned before, since the solver is independent of the interpreter, its improvements are
orthogonal to the core implementation of s(CASP).

The interpreter checks the call path before the evaluation of
user-defined predicates to prevent inconsistencies and infinite
loops~\marpleB (see~\ref{app:detect-loops}\supp).  The call path is a list
constructed with the calls, and the bindings of the variables in these
calls are automatically updated by Prolog.

\reditemize
\begin{itemize}
\item When a positive loops occurs, the interpreter fails only if the
  looping goal and its ancestor are equal (i.e.,
  \code{p(X):-$\ldots$,p(X)}). Termination properties are enhanced if
  a tabling system featuring variant calls or
  entailment~\cite{arias-ppdp2016} is used as implementation
  target, so that all programs with a finite grounding or with the
  constraint-compact property terminate. 
\item However, when the current call is equal to an already-proven
  ancestor, the evaluation succeeds to avoid its re-computation and to
  reduce the size of the justification tree.
\end{itemize}
\reditemize

\redbefsec
\subsection{Integration of  Constraint Solvers in s(CASP)}
\label{sec:integr-constr-solv}

Holzbaur's CLP($\mathds{Q}$)~\cite{holzbaur-clpqr} solver was integrated in the
current implementation of s(CASP).  Since the interpreter already
deals with the CLP($\not=$) constraint solver, %
only two details have to be taken in consideration:

\reditemize
\begin{itemize}
\item The compiler is extended to support CLP($\mathds{Q}$) relations
  $\{<, >, =, \geq, \leq, \neq\}$ during the construction of the dual
  program and the NMR rules.
\item Since it is not possible to decide at compile time whether
  equality will be called with CLP($\mathds{Q}$) or Herbrand variables, its dual
  \verb'\=' is extended to decide at run-time whether to call the
  CLP($\mathds{Q}$) solver or the disequality solver. 
\end{itemize}
\reditemize

Finally, to make integrating further constraint solvers easier, the
operations that the s(CASP) interpreter requires from the CLP($\mathds{Q}$)
solver are encapsulated in a single module that provides the interface
between the interpreter and the constraint solver.  Additional
constraint solvers only need to provide the same interface.

\begin{algorithm}[t]

  \emph{C-forall} receives the variable \code{V}, the
  callable goal \code{Goal}, and an initial constraint store, $C_i$
  ($i=1$).

  \code{V} starts unbound. 
   \tpl*[f]{\emph{The constraint store of \code{V} is empty, $C_{V.i} = \top$}}

  Execute \code{Goal} with $C_i$ as the current constraint store.
  \tpl*[f]{\emph{Its first answer constraint store is $A_{1}$}}

  \label{item:1}
  \uIf (\tpl*[f]{\emph{Check $A_{V.i}$, the domain of
      \code{V} in the answer constraint}}){the execution of \code{Goal} succeeds}{%

    \uIf (\tpl*[f]{\emph{There was no refinement in the domain of
        \code{V}}}){$A_{V.i} \equiv C_{V.i}$}{%

      \emph{C-forall} \textbf{succeeds} \tpl*[f]{\emph{\code{V} is not
        relevant for the success of \code{Goal}}}

    }

    \Else(\tpl*[f]{\emph{The domain of \code{V}
         has been restricted, $A_{V.i} \sqsubset C_{V.i}$}}) {%

       $C_{i+1} = C_i \land A_{\overline{V}.i} \land \lnot A_{V.i}$
       \tpl*[f]{\emph{Remove from \code{V} the elements for which
          \code{Goal} succeeds}}
       
      Return to step~\ref{item:1} and re-execute \code{Goal}
      under $C_{i+1}$

      \tpl*[f]{\emph{Check whether \code{Goal} is \code{true} for
          the rest of the elements of \code{V}}}

    }

  }
  
  \lElse(\tpl*[f]{\emph{There is a non-empty domain for which
      \code{Goal} is not true}}){\emph{C-forall} \textbf{fails}}

\caption{C-forall}
\label{alg:forall2}
\end{algorithm}

\redbefsec
\subsection{Extending \texttt{forall}\, for Constraints}
\label{sec:constr-forall}

Extending s(ASP) to programs with constraints requires a
generalization of \emph{forall} (Algorithm~\ref{alg:forall})
which we will call
\emph{C-forall} (Algorithm~\ref{alg:forall2}).  A successful evaluation of
\code{Goal} in s(CASP) returns, on backtracking, a (potentially
infinite) sequence of models and answer constraint stores
$A_1, A_2, \dots$.  Each $A_i$ relates variables and constants by
means of constraints and bindings (i.e., syntactical equality
constraints).  The execution of \code{forall(V,Goal)} is expected to
determine if \code{Goal} is true for all possible values of
\code{V} in its constraint domain.

In what follows we will use $\cV$ to denote the set variables 
in \code{Goal} that are not \code{V}: $vars$(\code{Goal}) =
$\{\V\} \cup \cV \land \V \not\in \cV$. 
The core idea is to iteratively narrow the store $C$ under which
\code{Goal} is executed by
selecting \textbf{one} answer $A$ and re-executing \code{Goal} under
the constraint store
$C \land A_{\cV} \land \lnot A_{\V}$, where $A_{\V}$ is
the projection of $A$ on \code{V} and $A_{\cV}$ is the
projection of $A$ on \code{$\cV$}.  The iterative execution
finishes with a positive or negative outcome.

\redmiditemize

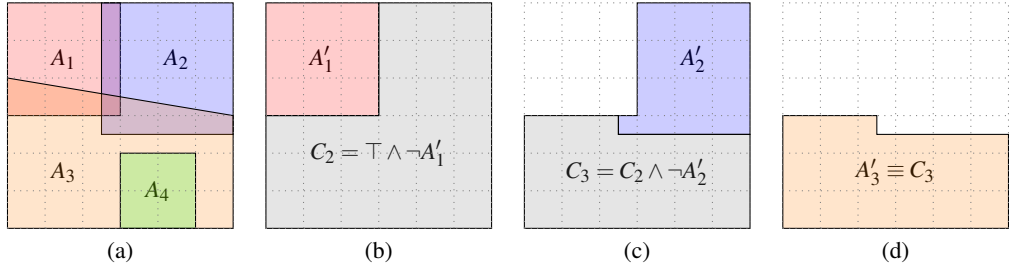
\begin{figure}
  \centering
  \begin{tabular}{cccc}
    \begin{tikzpicture}
      \node (a) at (.75,2.25) {$A_1$};
      \node (b) at (2.25,2.25) {$A_2$};
      \node (c) at (.75,.75) {$A_3$};
      \node (d) at (2,.5) {$A_4$};
      \draw[fill=red, fill opacity=.2] (0,1.5) -- (0,3) -- (1.5,3) --
      (1.5,1.5) -- cycle;

      \draw[fill=blue, fill opacity=.2] (1.25,3) -- (3,3) -- (3,1.25) -- (1.25,1.25) -- cycle;
      \draw[fill=orange, fill opacity=.2] (0,0) -- (0,2) -- (3,1.5) -- (3,0) -- cycle; 
      \draw[fill=green, fill opacity=.2] (1.5,0) -- (1.5,1) -- (2.5,1) -- (2.5,0) -- cycle; 
      \draw[color=gray, style=dotted] (0,0) 
      grid[xstep=.5cm, ystep=.5cm] (3cm,3cm);
    \end{tikzpicture}
    &
      \begin{tikzpicture}
        \node (a) at (.75,2.25) {$A'_1$};
        \node (b) at (1.5,1) {$C_2=\top \land \lnot A'_1$};
        \draw[fill=red, fill opacity=.2] (0,1.5) -- (0,3) -- (1.5,3) -- (1.5,1.5) -- cycle;
        \draw[fill=gray, fill opacity=.2] (0,0) -- (0,1.5) -- (1.5,1.5) -- (1.5,3) -- (3,3) -- (3,0) -- cycle;
        \draw[color=gray, style=dotted] (0,0) 
        grid[xstep=.5cm, ystep=.5cm] (3cm,3cm);
      \end{tikzpicture}
    &
      \begin{tikzpicture}
        \node (b) at (2.25,2.25) {$A'_2$};
        \node (c) at (1.5,.75) {$C_3=C_2 \land \lnot A'_2$};
        \draw[fill=blue, fill opacity=.2] (1.5,3) -- (3,3) -- (3,1.25) -- (1.25,1.25) -- (1.25,1.5) -- (1.5,1.5) -- cycle;
        \draw[fill=gray, fill opacity=.2] (0,0) -- (0,1.5) -- (1.25,1.5) -- (1.25,1.25) -- (3,1.25) -- (3,0) -- cycle;
        \draw[color=gray, style=dotted] (0,0) 
        grid[xstep=.5cm, ystep=.5cm] (3cm,3cm);
      \end{tikzpicture}
    &
      \begin{tikzpicture}
        \node (c) at (1.5,.75) {$A'_3 \equiv C_3$};
        \draw[fill=orange, fill opacity=.2] (0,0) -- (0,1.5) -- (1.25,1.5) -- (1.25,1.25) -- (3,1.25) -- (3,0) -- cycle;
        \draw[color=gray, style=dotted] (0,0) 
        grid[xstep=.5cm, ystep=.5cm] (3cm,3cm);
      \end{tikzpicture}
      \\ (a) & (b) & (c) & (d)
  \end{tabular}
  \caption{A \emph{C-forall} evaluation that succeeds.}
  \label{fig:forall-success}
\end{figure}

\begin{figure}
  \centering
  \begin{tabular}{cccc}
    \begin{tikzpicture}
      \node (a) at (.75,2.25) {$A_1$};
      \node (b) at (2.5,2.25) {$A_2$};
      \node (c) at (.75,.75) {$A_3$};
      \node (d) at (2,0.5) {$A_4$};
      \draw[fill=red, fill opacity=.2] (0,1.5) -- (0,3) -- (1.5,3) -- (1.5,1.5) -- cycle;
      \draw[fill=blue, fill opacity=.2] (2,3) -- (3,3) -- (3,1.25) -- (2,1.25) -- cycle;
      \draw[fill=orange, fill opacity=.2] (0,0) -- (0,2) -- (3,1.5) -- (3,0) -- cycle; 
      \draw[fill=green, fill opacity=.2] (1.5,0) -- (1.5,1) -- (2.5,1) -- (2.5,0) -- cycle; 
      \draw[color=gray, style=dotted] (0,0) 
      grid[xstep=.5cm, ystep=.5cm] (3cm,3cm);
    \end{tikzpicture}
    &
      \begin{tikzpicture}
        \node (a) at (.75,2.25) {$A'_1$};
        \node (b) at (1.5,1) {$C_2=\top \land \lnot A'_1$};
        \draw[fill=red, fill opacity=.2] (0,1.5) -- (0,3) -- (1.5,3) -- (1.5,1.5) -- cycle;
        \draw[fill=gray, fill opacity=.2] (0,0) -- (0,1.5) -- (1.5,1.5) -- (1.5,3) -- (3,3) -- (3,0) -- cycle;
        \draw[color=gray, style=dotted] (0,0) 
        grid[xstep=.5cm, ystep=.5cm] (3cm,3cm);
      \end{tikzpicture}
    &
      \begin{tikzpicture}
        \node (b) at (2.5,2.25) {$A'_2$};
        \node (c) at (1.5,.75) {$C_3=C_2 \land \lnot A'_2$};
        \draw[fill=blue, fill opacity=.2] (2,3) -- (3,3) -- (3,1.25) -- (2,1.25) -- cycle;
        \draw[fill=gray, fill opacity=.2] (0,0) -- (0,1.5) -- (1.5,1.5) -- (1.5,3) -- (2,3) -- (2,1.25) -- (3,1.25) -- (3,0) -- cycle;
        \draw[color=gray, style=dotted] (0,0) 
        grid[xstep=.5cm, ystep=.5cm] (3cm,3cm);
      \end{tikzpicture}
    &
      \begin{tikzpicture}
        \node (c) at (1.5,.75) {$A'_3$};
        \node (c) at (1.75,2.25) {$C_4$};
        \draw[fill=orange, fill opacity=.2] (0,0) -- (0,1.5) -- (1.5,1.5) -- (1.5,1.75) -- (2,1.65) -- (2,1.25) -- (3,1.25) -- (3,0) -- cycle;
        \draw[fill=gray, fill opacity=.2] (1.5,3) -- (1.5,1.75) -- (2,1.65) -- (2,3) -- cycle;
        \draw[color=gray, style=dotted] (0,0) 
        grid[xstep=.5cm, ystep=.5cm] (3cm,3cm);
      \end{tikzpicture}
      \\ (a) & (b) & (c) & (d)
  \end{tabular}
  \caption{A \emph{C-forall} evaluation that fails.}
  \label{fig:forall-failure}
\redbefsec
\end{figure}
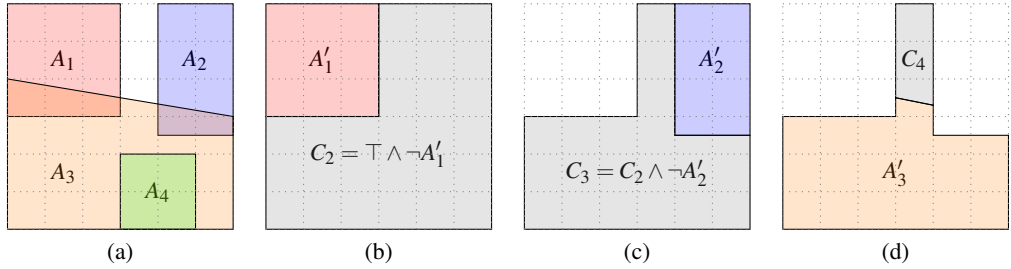

\begin{example}[\emph{C-forall} terminates with success]
  Figure~\ref{fig:forall-success} shows an example where the answers
  $A_1,\ldots,A_4$ to \code{Goal} cover the whole domain, represented
  by the square.  Therefore, \emph{C-forall} should succeed.  The
  answer constraints that the program can generate are depicted on
  picture (a).  For simplicity in the pictures, we will assume that
  the answers $A_i$ only restrict the domain of $\V$, so it will not
  be necessary to deal with \V and \cV separately since $A_{\cV}$ will
  always be empty, and therefore $A_{\V.i}=A_i$.  Picture (b) shows
  the result of the first iteration of \emph{C-forall} starting with
  $C_1 = \top$: answer $A_1$ is more restrictive than $C_1$ and
  therefore $C_2 = C_1 \land \lnot A_1$ (in grey) is constructed.
  Picture (c) shows the result of the second iteration: the domain is
  further reduced.  Finally, in picture (d) the algorithm finishes
  successfully because $A_3 \equiv C_3$, i.e., $A_3$ covers the
  remaining domain.  Note that we did not need to generate $A_4$.
\end{example}

\noindent
\paragraph{\textbf{Termination for an infinite number of answer sets}}

The previous example points to a nice property: even if there were an
infinite number of answer sets to \code{Goal}, as long as a finite
subset of them covers the domain of \code{V} and this subset can be
finitely enumerated by the program, the algorithm will finish.  This
is always true for constraint-compact domains, such as disequality
over a finite set of constants or the gap-order
constraints~\cite{revesz1993closed}.  Note that this happens as well
in the next example, where \emph{C-forall} fails.

\begin{example}[\emph{C-forall} terminates with failure]
  Figure~\ref{fig:forall-failure} shows an example where the answer
  constraints do not cover the domain and therefore \emph{C-forall}
  ought to fail.  Again, we assume that the answers $A_i$
  only restrict the domain of $\V$. %
  Picture (a) depicts the answer constraints that \code{Goal} can
  generate.  Note the gap in the domain not covered by the answers.
  Pictures (b) to (d) proceed as in the previous example.  Picture (d)
  shows the final step of the algorithm: the execution of \code{Goal}
  under the store $C_4=C_3 \land \lnot A'_3$ fails because the
  solution $A_4$ of \code{Goal} does not have any element in common
  with $C_4$, and then \emph{C-forall} also fails.
\end{example}


\begin{figure}
\begin{lstlisting}[style=MyProlog]
forall(V, Goal) :-
   empty_store(Store),             % V has no attached constraints
   eval_forall(V, Goal, [Store]).  % start the evaluation of Goal
eval_forall(_, _, []).             % it's done, forall succeeds   
eval_forall(V,Goal,[Store|Sts]):-
   copy(V, Goal, NV, NGoal),       % copy to keep V unbound
   apply(NV, V, Store),            % add the constraint to NV
   once(NGoal),                    % if fails, the forall fails
   dump(NV, V, AnsSt),             % project the answer store
   &(   equal(AnsSt, Store)         % if there is no refinement in NV
   ->  true                        % then, it's done, continue
   ;   dual(AnsSt, AnsDs),         % else, the answer's dual/duals
       add(AnsDs, Store, NSt),     % is/are added to Store
       eval_forall(V, Goal, NSt)   % to evaluate Goal
   &),
   eval_forall(V, Goal, Sts).      % continue the evaluation
\end{lstlisting}
  \caption{Code of the predicate \code{forall/2} implemented in s(CASP).}
  \label{fig:forall}
\redbefsec
\end{figure}

\noindent
Figure~\ref{fig:forall} shows a sketch of the code that implements
\emph{C-forall} in the s(CASP) interpreter, written in Prolog/CLP.  In
this setting, \code{Goal} carries the constraint stores $C_i$ and the
answer stores $A_i$ implicitly in its execution environment.  We know
that the interpreter will call \code{forall(V,Goal)} with a fresh,
unconstrained \code{V}, because the executed code is generated by the
s(CASP) compiler.  Therefore, the projection of $C_1$ onto \code{V} is
an empty constraint store, which we introduce explicitly to start the
computation.

The call %
\code{copy(V,Goal,NV,NGoal)} copies \code{Goal} in \code{NGoal}
sharing only $\cV$, while \code{V} is substituted in \code{NGoal} by a
fresh variable, \code{NV}.
In the main body of \code{eval_forall/3}, \code{Store} always refers
to \V, while \code{NGoal} does not contain \V, but \code{NV}.  The
call \code{apply(NV,V,Store)} takes the object \code{Store} and makes
it part of the global store but substituting \V for \code{NV} so that
the execution of \code{NGoal} can further constrain \code{NV} while \V
remains untouched.  Note, however, that in the first iteration,
\code{NV} will always remain unconstrained, since the constraint store
that \code{apply(NV,V,Store)} applies to it is empty
($C_{\V.1}=\top$).  However, in the following iterations, \code{Store}
will contain the successive constraint stores $C_{\V.i+1}$.

When \code{once(NGoal)} succeeds, the constraint store
$C_{i} \land A_{\cV.i}$ is implicit in the binding of $\cV$.
Therefore, the execution of \code{eval_forall(V,Goal,Store)} carries
this constraint store implicitly because \code{Goal} and \code{NGoal}
share \cV.
Finally, the predicate \code{dump(NV,V,AnsSt)} projects the constraint
store after the execution of \code{NGoal} on \code{NV}, rewrites this
projection to substitute \code{NV} for \V, and leaves the final result
in \code{AnsSt}, generating $A_{\V.i}$.  Note that, in some sense, it
is transferring constraints in the opposite direction to what
\code{dump/3} did before.
If the call \code{equal(AnsSt, Store)} succeeds, it means that
$A_{\V.i} \equiv C_{\V.i}$ and therefore the \code{forall} succeeds (for
the branch that was being explored, see below).

Otherwise, we have to negate the projection of the answer onto
\code{V}, i.e., construct $\lnot A_{\V.i}$.  The negation of a
conjunction generates a disjunction of constraints and most constraint
solvers cannot handle disjunctions natively.  Therefore, the predicate
\code{dual(AnsSt,AnsDs)} returns in \code{AnsDs} a list with the
components $\lnot A_{\V.i.j}$ of this disjunction,
$j = 1, 2, \ldots, \mathit{length}(\mathtt{AnsDs})$.
Then, \code{add(AnsDs,Store,NSt)} returns in \code{NSt} a list of
stores, each of which is the conjunction of \code{Store} with one of
the components of the disjunction in \code{AnsDs}, i.e., a list of
$C_{\V.i} \land \lnot A_{\V.i.j}$, for a fixed $i$.  There may be
cases where this conjunction is inconsistent; \code{add/3} captures
them and returns only the components which are consistent.  Note that
if a conjunction $C_{\V.i} \land \lnot A_{\V.i.j}$ is inconsistent, it
means that $\lnot A_{\V.i.j}$ has already been (successfully) checked.

Each of the resulting constraint stores will be re-evaluated by
\code{eval_forall/3}, where \code{apply/3} will apply them to a new
variable \code{NV}, in order to complete the implicit construction of
$C_{i+1}$ before the execution of \code{once(NGoal)}.
\code{forall/2} finishes with success when there are no pending
constraint stores to be processed (line 4).

\begin{example}[\texttt{C-forall} execution negating a constraint conjunction]
  Given the program below, consider the evaluation of
  \code{forall(A, p(A))}:
  \begin{center}
  \begin{minipage}{.9\linewidth}
  \redbeflst
\begin{multicols}{2}
\begin{lstlisting}[style=MyProlog]
p(X) :- X.>=.0, X.=<.5.
p(X) :- X.>.1.
p(X) :- X.<.3.
p(X) :- X.<.1.
\end{lstlisting}
  \end{multicols}
\end{minipage}
\end{center}

\medskip

In the first iteration $C_1=\top$. The first answer is
$A_1=\{X \geq 0 \land X \leq 5\}$, which is more restrictive than $C_1$, so
we compute $\lnot A_1=\{X < 0 \lor X > 5\}$.
First, \code{p/1} is evaluated with $C_{2.a}=\{\top \land X < 0\}$
obtaining $A_{2.a}=\{X < 0\}$ using the third clause. Since
$A_{2.a} \equiv C_{2.a}$, we are done with $C_{2.a}$. But we also have
to evaluate \code{p/1} with $C_{2.b}=\{\top \land X > 5\}$.  Using the second
clause, $A_{2.b}=\{X > 5\}$ is obtained and since
$A_{2.b} \equiv C_{2.b}$, the evaluation succeeds.
\end{example}

\redbefsec
\section{Examples and Evaluation}
\label{sec:examples}

The expressiveness of s(CASP) allows the programmer to write programs /
queries that cannot be written in [C]ASP without resorting to a
complex, unnatural encoding. Additionally, the answers given by
s(CASP) are also more expressive than those given by ASP. This arises
from several points:

\reditemize
\begin{itemize}
\item s(CASP) inherits from s(ASP) the use of unbound variables during
  the execution and in the answers.  This makes it possible to
  express constraints more compactly and naturally (e.g., ranges of
  distances can be written using constraints)
\item s(CASP) can use structures / functors directly, thereby avoiding
  the need to encode them unnaturally (e.g., giving numbers to Hanoi
  movements to represent what in a list is implicit in the sequence of its
  elements).
\item The constraints and the goal-directed evaluation strategy of s(CASP) makes
  it possible to use direct algorithms and to reduce
  the search space (e.g., by putting bounds on a path's length).
\end{itemize}
\reditemize

\redbefsec
\subsection{Stream Data Reasoning}
\label{sec:stream-reasoning}

Let us assume that we deal with data streams, some of whose items may
be contradictory~\cite{arias-dc-iclp2016}.  Moreover, different data
sources may have a different degree of trustworthiness which we use
to prefer a given data item in case of inconsistency.
Let us assume that \code{p(X)} and \code{q(X)} are contradictory
and we receive \code{p(X)} from source $S_1$  and \code{q(a)} from source
$S_2$.
We may decide, depending on how reliable are $S_1$ and $S_2$, that:
(i) \code{p(X)} is true because $S_1$ is more reliable than $S_2$;
(ii) \code{q(a)} is true since $S_2$ is more reliable than $S_1$, and
for any \code{X} different from \code{a} (i.e., \code{X\=a}),
\code{p(X)} is also true; (iii) or, if both sources are equally
reliable, them we have (at least) two different models: one where
\code{q(a)} is true and another where \code{p(X)} is true.

\begin{figure}
  \redbefsec
  \begin{minipage}{.9\linewidth}
\begin{multicols}{2}
\begin{lstlisting}[style=MyProlog]
valid_stream(P,Data) :- 
     stream(P,Data), 
     not cancelled(P, Data).

cancelled(P, Data) :- 
     higher_prio(P1, P), 
     stream(P1, Data1), 
     incompt(Data, Data1).

higher_prio(PHi, PLo) :- 
     PHi.>.PLo.
incompt(p(X), q(X)).
incompt(q(X), p(X)).

stream(1,p(X)).   
stream(2,q(a)).
stream(2,q(b)).   
stream(3,p(a)).
\end{lstlisting}    
  \end{multicols}
\end{minipage}
\caption{Code of the stream reasoner.}
\label{fig:stream_code}
\redbefsec
\end{figure}

Figure~\ref{fig:stream_code} shows the code for a stream reasoner
using s(CASP).  Data items are represented as
\code{stream(Priority, Data)}, where \code{Priority} tells us the
degree of confidence in \code{Data};
\code{higher_prio(PHi,PLo)} hides how priorities
are encoded in the data (in this case, the higher the priority, the more level of
confidence);
and \code{incompt/2} determines which data items are
contradictory (in this case, \code{p(X)} and \code{q(X)}).
Note that \code{p(X)} (for \textbf{all}
\code{X}) has less
confidence than \code{q(a)} and \code{q(b)}, but \code{p(a)} is an exception, as it
has more confidence than \code{q(a)} or \code{q(b)}.
Lines 1-8, alone, define
the reasoner rules: \code{valid_stream/2} states that a data stream
is valid if it is \emph{not cancelled} by another contradictory data stream
with more confidence.

The confidence relationship uses constraints, instead of being checked
afterwards.  \emph{C-forall}, introduced by the compiler in the
dual program (\ref{app:appendix-stream}.1\supp),
will check its consistency.  For the query %
\code{?- valid_stream(Pr,Data)}, it returns:
\code{\{Pr=1, Data=p(A), A\=a, A\=b\}} \mbox{because} \code{q(a)} and
\code{q(b)} are more reliable than \code{p(X)};
\code{\{Pr=2, Data=q(b)\}}; and \code{\{Pr=3, Data=p(a)\}}. The
justification tree and the model are in \ref{app:appendix-stream}.2\supp.

The constraints and the goal-directed strategy of s(CASP) make it
possible to resolve queries without evaluating the whole stream
database.
For example, the rule \code{incompt(p(X),q(X))} does not have to be
grounded w.r.t.\ the stream database, and if timestamps were used as
trustworthiness measure,
for a query such as \code{?-T.>.10,valid_stream(T,p(A))} the reasoner
would validate streams
received after \code{T=10} regardless how long they extend in the
past.

\redbefsec
\subsection{Yale Shooting Scenario} 
\label{sec:yale-shoot-scen}

\begin{figure}
  \hspace*{-.05\linewidth}
  \begin{minipage}{.95\linewidth}
\begin{multicols}{2}
\begin{lstlisting}[style=MyProlog, basewidth=.53em]
duration(load,25).
duration(shoot,5).
duration(wait,36).
spoiled(Armed) :- Armed #> 35.
prohibited(shoot,T) :- T #< 35.

holds(0,St,[]) :- init(St).
holds(F_Time, F_St, [Act|As]) :-
   F_Time #> 0,
   F_Time #= P_Time + Duration,
   duration(Act, Duration),
   not prohibited(Act, F_Time),
   trans(Act, P_St, F_St),
   holds(P_Time, P_St, As).
init(st(alive,unloaded,0)).

trans(load, st(alive,_,_),
            st(alive,loaded,0)).
trans(wait, st(alive,Gun,P_Ar), 
            st(alive,Gun,F_Ar)) :-
   F_Ar #= P_Ar + Duration,
   duration(wait,Duration). 
trans(shoot, st(alive,loaded,Armed), 
             st(dead,unloaded,0)) :-
   not spoiled(Armed).
trans(shoot, st(alive,loaded,Armed), 
             st(alive,unloaded,0)) :-
   spoiled(Armed).
\end{lstlisting}
\end{multicols}
\end{minipage}
\caption{s(CASP) code for the Yale Shooting problem.}
\label{fig:yale-short}
\redbefsec
\end{figure}

In the spoiling Yale shooting scenario~\cite{janhunen2017clingo},
there is a gun and three possible actions: \emph{load}, \emph{shoot},
and \emph{wait}.
If we load the gun and shoot within 35 minutes, the turkey is killed.
Otherwise, the gun powder is spoiled. The executable plan must ensure
that we kill the turkey within 100 minutes, assuming that we are not
allowed to shoot in the first 35 minutes.

The ASP + constraint code, in~\cite{janhunen2017clingo} and
\ref{app:yale-example}.1\supp, uses \emph{clingo[DL/LP]}, an ASP
incremental solver 
extended for constraints. The program is parametric w.r.t.\ the step
counter $n$, used by the solver to iteratively invoke the program with
the expected length of the plan.
In each iteration, the solver increases $n$, grounds the
program with this value (which, in this example, specializes it for a
plan of exactly $n$ actions) and solves it.
The execution returns two plans for $n=3$:
\code{\{do(wait,1), do(load,2), do(shoot,3)\}} and
\code{\{do(load,1), do(load,2), do(shoot,3)\}}.

The s(CASP) code (Figure~\ref{fig:yale-short}) does not need a
counter. The query %
\code{?-T.<.100, holds(T,st(dead,_,_),Actions)}, sets an upper bound
to
the  duration \code{T} of the plan, and returns in
\code{Actions}  the plan with the  actions in reverse chronological
order: %
\code{\{T=55, Actions=[shoot, load, load]\}}, %
\code{\{T=66, Actions=[shoot, load, wait]\}}, %
\code{\{T=80, Actions=[shoot, load, load, load]\}}, %
\code{\{T=91, Actions=[shoot, load, load, wait]\}}, %
\code{\{T=91, Actions=[shoot, load, wait, load]\}}, %
\code{\{T=96, Actions=[shoot, load, shoot, wait, load]\}}.

\redbefsec
\subsection{The Traveling Salesman Problem (TSP)}
\label{sec:trav-salesman}

Let us consider a variant of the traveling salesman problem (visiting
every city in a country only once, starting and ending in the same
city, and moving between cities using the existing connections) where
we want to find out only the Hamiltonian cycles whose length is less
than a given upper bound.
Solutions for this problem, with comparable performance, using ASP and CLP($\mathit{FD}$)
appear in~\cite{dovier2005comparison} (also available at~\ref{app:trav-salesm-probl}.1 and~\ref{subapp:trav-salesman-fd}\supp).  The
ASP encoding is more compact, even if the CLP($\mathit{FD}$) version uses the non-trivial
library predicate \mbox{\code{circuit/1}}, which does the bulk of the work.
We will show that s(CASP) is more expressive also in this problem.

Finding the (bounded) path length in ASP requires using a specific,
ad-hoc builtin that accesses the literals in a model and calls it from
within a global constraint.  Using
\emph{clasp}~\cite{holldobler2014answer}, it would be as follows:

 \redbeflst

\begin{center}
\begin{minipage}{.95\linewidth}
\begin{lstlisting}[style=MyASP]
cycle_length(N) :- N = #sum [cycle(X,Y) : distance(X, Y, C) = C].
:- cycle_length(N), N >= 10.      % Cycles whose length is less than 10
\end{lstlisting}
\end{minipage}
\end{center}

\medskip

\noindent
where \code{#sum} is a builtin aggregate operator that here is used to
add the distances between nodes in some Hamiltonian cycle.

\begin{figure}
\begin{multicols}{2}
\begin{lstlisting}[style=MyProlog]
% Every node must be reachable.
:- node(U), not reachable(U).
reachable(a) :- cycle(V,a).
reachable(V) :- cycle(U,V), 
                reachable(U).

% Only one edge to each node.
:- cycle(U,W), cycle(V,W), U\=V.

% Only one edge from each node.
cycle(U,V) :-
  edge(U,V), not other(U,V).
other(U,V) :-
  node(U), node(V), node(W),
  edge(U,W), V\=W, cycle(U,W).

travel_path(S,Ln,Cycle) :- 
  path(S,S,S,Ln,[],Cycle).
path(_,X,Y,D,Ps,[X,[D],Y|Ps]) :-
  cycle_dist(X,Y,D).
path(S,X,Y, D, Ps,Cs) :-
  D.=.D1 + D2,
  cycle_dist(Z,Y,D1), Z\=S,
  path(S,X,Z,D2,[[D1],Y|Ps],Cs).

edge(X,Y) :- distance(X,Y,D).
cycle_dist(U,V,D) :- 
  cycle(U,V), distance(U,V,D).

node(a).          node(b).
node(c).          node(d).

distance(b,c,31/10).
distance(c,d,L):- 
    L.>.8, L.<.21/2.
distance(d,a,1).
distance(a,b,1).  
distance(a,d,1).
distance(c,a,1).  
distance(d,b,1).
\end{lstlisting}
\end{multicols}
\caption{Code for the Traveling Salesman problem.}
\label{fig:travel-salesman}
\redbefsec
\end{figure}

The s(CASP) code in Figure~\ref{fig:travel-salesman} solves this TSP
variant by modeling the Hamiltonian cycle in a manner similar to ASP
and using a recursive predicate, \code{travel_path(S,Ln,Cycle)}, that
returns in \code{Cycle} the list of nodes in the circuit (with the
distance between every pair of nodes also in the list), starting at
node \code{S}, and the total length of the circuit in \code{Ln}.

This example highlights the marriage between ASP
encoding (to define models of the Hamiltonian cycle using the
\code{cycle/2} literal) and traditional
CLP (which uses the available \mbox{\code{cycle/2}} literals to construct
paths and return their lengths).
Note as well that we can define node distances as intervals (line 35)
using a dense domain (rationals, in this case).  This would not be
straightforward (or even feasible) if only CLP($\mathit{FD}$) was available:
while CLP($\mathit{FD}$) can encode CLP($\mathds{Q}$), the resulting program would be cumbersome
to maintain and much slower than the CLP($\mathds{Q}$) version, since
Gaussian elimination has to be replaced by enumeration, which actually
compromises completeness (and, in the limit, termination).
Additionally, in our proposal, constraints can appear in bindings and
as part of the model.
For example, the query \code{?-D.<.10, travel_path(b,D,Cycle)}
returns
the model \code{\{D=61/10, Cycle=[b, [31/10], c, [1], a, [1], d, [1], b]\}}.
For reference, \ref{app:trav-salesm-probl}.3\supp\ shows the complete output.

\redbefsec
\subsection{Towers of Hanoi}
\label{sec:eval-towers-hanoi}

We will not explain this problem here as it is widely known.
Let us just remind the reader that solving the puzzle with three
towers (the standard setup) and $n$ disks requires at least $2^n - 1$
movements.

\begin{figure}
  \hspace*{-.1\linewidth}
  \begin{minipage}{.90\linewidth}
\begin{multicols}{2}
\begin{lstlisting}[style=MyProlog]
hanoi(N,T):- 
   move_(N,0,T,a,b,c).
move_(N,Ti,Tf,Pi,Pf,Px) :-
   N.>.1, N1.=.N - 1,
   move_(N1,Ti,T1,Pi,Px,Pf),
   move_( 1,T1,T2,Pi,Pf,Px),
   move_(N1,T2,Tf,Px,Pf,Pi).
move_(1,Ti,Tf,Pi,Pf,_) :-
   Tf.=.Ti + 1,
   move(Pi,Pf,Tf).
move(Pi,Pf,T):- not negmove(Pi,Pf,T).
negmove(Pi,Pf,T):- not move(Pi,Pf,T).

#show move/3. %s(CASP) directive
\end{lstlisting}
\end{multicols}
\end{minipage}
\caption{s(CASP) code for the Towers of Hanoi.}
\label{fig:hanoi}
\redbefsec
\end{figure}

Known ASP encodings, for a \emph{ standard} solver, set a bound to the
number of moves that can be done, as proposed in~\cite{gebser2008user}
(available for the reader's convenience at~\ref{app:towers-hanoi}.1\supp,
for \code{7} disks and up to \code{127} movements) or for an
\emph{incremental} solver, increasing the number $n$ of allowed
movements (from the \emph{clingo 5.2.0} distribution,
also available at \ref{app:towers-hanoi}.2\supp).

s(CASP)'s top down approach can use a CLP-like control strategy to
implement the well-known Towers of Hanoi algorithm
(Figure~\ref{fig:hanoi}).  Predicate \code{hanoi(N,T)} receives in
\code{N} the number of disks and returns in \code{T} the number of
movements needed to solve the puzzle. The resulting partial stable
model will contain all the movements and the time in which they have
to be performed.
For reference, \ref{app:towers-hanoi}.3\supp\ shows the partial stable
model for \code{?-hanoi(7,T)}.

\begin{table} 
  \redbefwrap\normalsize
 \setlength{\tabcolsep}{1.75em}
  \begin{tabular}{@{\extracolsep{0pt}}lrrr}
    & \textbf{s(CASP)} & \textbf{clingo 5.2.0} & \textbf{clingo 5.2.0} \\ 
    &  & \multicolumn{1}{c}{\textit{standard}} & \multicolumn{1}{c}{\textit{incremental}} \\ \cline{1-4}
    $n=7$ & \textbf{479}   & 3,651   &  9,885    \\
    $n=8$ & \textbf{1,499} & 54,104  &  174,224  \\ 
    $n=9$ & \textbf{5,178} & 191,267 & $>$ 5 min \\ \cline{1-4}
  \end{tabular}
  \caption{Run time (ms) comparison for the Towers of Hanoi with $n$
    disks.}
  \label{tab:eval-hanoi}
\end{table}

Table~\ref{tab:eval-hanoi} compares execution time (in milliseconds)
needed to solve the Towers of Hanoi with \code{n} disks by s(CASP) and
\emph{clingo 5.2.0} with the \emph{standard} and \emph{incremental}
encodings.  s(CASP) is orders of magnitude faster than both clingo
variants because it does not have to generate and test all the
possible plans; instead, as mentioned before, it computes directly the
smallest solution to the problem. The standard variant is less
interesting than s(CASP)'s, as it does not return the minimal number
of moves --- it merely checks if the problem can be solved in a given
number of moves.  The incremental variant is by far the slowest,
because the program is iteratively checked with an increasing number
of moves until it can be solved.

\redbefsec
\section{Conclusion and Future Work}
\label{sec:concl-future-work}

We have reported on the design and implementation of s(CASP), a
top-down system to evaluate constraint answer set programs, based on
s(ASP).  Its ability to express non-monotonic programs \emph{\`a la} ASP is
coupled with the possibility of expressing control in a way similar to
traditional logic programming --- and, in fact, a single program can
use both approaches simultaneously, achieving the best of both worlds.
We have also reported a very substantial performance increase w.r.t.\
the original s(ASP) implementation.  Thanks to the possibility of
writing pieces of code with control in mind, it can also beat
state-of-the-art ASP systems in certain programs.

The implementation can still be improved substantially, as pointed out
in the paper, and in particular we want to work on using analysis to
optimize the compilation of non-monotonic check rules, being able to
interleave their execution with the top-down strategy to discard
models as soon as they are shown inconsistent, improve the disequality
constraint solver to handle the pending cases, use dependency
analysis to improve the generation of the dual programs, and apply partial
evaluation and better compilation techniques to remove (part of) the overhead
brought about by the interpreting approach.

\redbefsec
\bibliographystyle{acmtrans}

\newpage


\appendix

\normalsize
\section{s(CASP) interpreter}
\label{app:scasp-interpreter}

The next figure shows a sketch of the s(CASP) interpreter's code
implemented in Ciao Prolog.

\begin{multicols}{2}
\begin{lstlisting}[style=MyProlog, basewidth=.51em]
??(Query) :-
  solve(Query,[],Mid),
  solve_goal(nmr_check,Mid,Just),
  print_just_model(Just).

solve([],In,['\$success'|In]).
solve([Goal|Gs],In,Out) :-
  solve_goal(Goal,In,Mid),
  solve(Gs,Mid,Out).

solve_goal(Goal,In,Out) :-
  user_defined(Goal),!,
  check_loops(Goal,In,Out).
solve_goal(Goal,In,Out) :-
  Goal=forall(Var,G),!,
  forall(V,G,In,Out).
solve_goal(Goal,In,Out) :-
  call(Goal),
  Out=['\$success',Goal|In].

check_loops(Goal,In Out) :-
  type_loop(Goal,In,Loop),
  solve_loop(Loop,Goal,In,Out).

solve_loop(odd,_,_,_) :- fail.
solve_loop(pos,_,_,_) :- fail.
solve_loop(eve,G,In,[chs(G)|In]).
solve_loop(pro,G,In,[pro(G)|In]).
solve_loop(cont,G,In,Out) :-
  pr_rule(G, Body),
  solve(Body,[G|In],Out).

forall(V,Goal,In,Out) :-
  empty_store(Store),            
  eval_forall(V,Goal,[Store],In,Out).
eval_forall(_,_,[],In,In).
eval_forall(V,Goal,[Store|Sts],In,Out) :-
  copy(V,Goal, NV,NGoal),
  apply(NV, V,Store),
  solve([NGoal],In,['\$success'|Out_1]),            
  dump(NV, V,AnsSt),             
  &(  equal(AnsSt,Store)         
  -> Out_2 = Out_1          
  ;  dual(AnsSt,AnsDs),         
     add(AnsDs,Store,NSt),  
     eval_forall(V,Goal,NSt,Out_1,Out_2)        
  &),
  eval_forall(V,Goal,Sts,Out_2,Out).           
\end{lstlisting}
\end{multicols}

\section{Handling Loops}
\label{app:detect-loops}

Top-down evaluations may enter loops.  Several techniques, notably
tabling,
have been used to enhance the termination properties of LP systems.
This is more relevant in s(ASP) because the presence of
negation introduces new types of loops:

\begin{itemize}
\item \textbf{Odd loop over negation}: it occurs when a cycle in the
  call graph contains an odd number of intervening negations. These
  loops are important because they place global
  constraints 
  which restrict which literals can appear in a model. s(ASP) ensures
  that these global constraints are satisfied by introducing
  \emph{non monotonic rules}
  (Section~\ref{sec:non-monotonic-rules}). The odd loops are detected
  with a static analysis of the call graph checking the number of
  negations between recursive calls.

  \begin{example}
    The rules below, which are equivalent if \code{p/0} can not be
    added to the model by another rule, generate odd loops and force
    the stable model to satisfy $\lnot\ q(a)$.
    \begin{center}
      \begin{minipage}{.9\linewidth}
        \redbeflst    \begin{multicols}{2}
\begin{lstlisting}[style=MyProlog]
p :- q(a), not p.
:- q(a).
\end{lstlisting}
        \end{multicols}
      \end{minipage}
    \end{center}
  \end{example}

  \paragraph{\textbf{Run-time check of odd loops}}
  When, during the execution, a call unifies with its negation in the
  call path, the execution fails and backtracks.  Had it succeeded, it
  would have introduced a contradiction, and therefore the resulting
  partial stable model would have been discarded.
 
\item \textbf{Even loop over negation}: This happens when a call
  unifies with an ancestor in the call path and there is an even,
  non-zero, number of intervening negated calls between them. In this
  case, the execution succeeds assuming that the recursive call
  (partially) supports the \emph{negation} of those calls. The spirit
  underlying this assumption is similar to coinductive SLD
  resolution~\cite{gupta2007coinductive}, used to compute the greatest
  fixpoint of a program. Note that the Gelfond--Lifschitz method
  computes the fixpoint of the residual program, which is between the
  least fixpoint (computed by a top-down execution) and the greatest
  fixpoint. This assumption is safe because in cases where the
  evaluation tries to make this recursive call \emph{true}, the
  \emph{non monotonic rules} and the run-time detection of odd loops
  will discard the model.

  \begin{example}
    Consider the next program (with its dual) and the query
    \code{?-p(a)}.
  \begin{center}
    \begin{minipage}{.9\linewidth}
      \redbeflst
\begin{multicols}{2}
\begin{lstlisting}[style=MyProlog]
p(X) :- not q(X).                

q(X) :- not p(X).      
q(b).

not p(X) :- not p1(X).           
not p1(X) :- q(X).               
not q(X) :- not q1(X), not q2(X). 
not q1(X) :- p(X).
not q2(X) :- X\=b.
\end{lstlisting}
      \end{multicols}
    \end{minipage}
  \end{center}

  \medskip
  
  The call path %
  \code{p(a)$\leadsto$not q(a)$\leadsto$not q1(a)$\leadsto$p(a)}, %
  resulting from the query, shows that assuming \code{p(a)} we support
  both negated calls (i.e., \code{not q(a)} and %
  \code{not q1(a)}). Note that \code{not q(a)} is only partially
  supported because it succeeds only if also \code{not q2(X)}
  succeeds. Therefore, while the query \code{?-p(a)} succeeds, the
  query \code{?-p(b)} fails.%
\end{example}

\item \textbf{Positive loops}: when a call unifies with an ancestor in
  the call path and there are no intervening negative calls between
  them, the original s(ASP) fails to avoid infinite loops.  However,
  this behaviour compromises completeness and soundness. We work
  around this by checking that the call and its ancestor are equal
  (Section~\ref{sec:interpreter}).

  \begin{example}
    The next program generates infinitely many answers to the query
    \code{?-nat(X)}.
    \begin{center}
      \hspace*{-2em}
    \begin{minipage}{.7\linewidth}
      \redbeflst
\begin{multicols}{2}
\begin{lstlisting}[style=MyProlog]
nat(0).
nat(X) :- nat(Y), X is Y+1.
\end{lstlisting}
      \end{multicols}
    \end{minipage}
  \end{center}

  \medskip

  However, if it fails, when the recursive call \code{nat(Y)}
  unifies with its ancestor in the call path (i.e., the query), it
  loses completeness as it only returns the answer \code{X=0}, and
  therefore, due to the presence of negation, it also loses soundness.
\end{example}

\end{itemize}

\section{Stream Data Reasoning Example}
\label{app:appendix-stream}

\subsection{s(CASP) encoding of stream.pl}

The next figure shows the code of stream.pl with the dual program and the NMR generated by the extended compiler of s(CASP).

\begin{multicols}{2}
\begin{lstlisting}[style=MyProlog, basewidth=.52em]
valid_stream(P,Data) :-
    stream(P,Data),
    not cancelled(P,Data).

cancelled(P,Data) :-
    higher_prio(P1,P),
    stream(P1,Data1),
    incompt(Data,Data1).

higher_prio(PHi,PLo) :-
    PHi.>.PLo.

incompt(p(X),q(X)).
incompt(q(X),p(X)).

stream(1,p(X)).
stream(2,q(a)).
stream(2,q(b)).
stream(3,p(a)).

not incompt1(A,_,X) :-
    A\=p(X).
not incompt1(A,B,X) :-
    A=p(X),
    B\=q(X).

not incompt1(A,B) :-
    forall(X,not incompt1(A,B,X)).

not incompt2(A,_,X) :-
    A\=q(X).
not incompt2(A,B,X) :-
    A=q(X),
    B\=p(X).

not incompt2(A,B) :-
    forall(X,not incompt2(A,B,X)).

not incompt(A,B) :-
    not incompt1(A,B),
    not incompt2(A,B).

not higher_prio1(PHi,PLo) :-
    PHi.=<.PLo.

not higher_prio(A,B) :-
    not higher_prio1(A,B).

not cancelled1(P,_,P1,_) :-
    not higher_prio(P1,P).
not cancelled1(P,_,P1,Data1) :-
    higher_prio(P1,P),
    not stream(P1,Data1).
not cancelled1(P,Data,P1,Data1) :-
    higher_prio(P1,P),
    stream(P1,Data1),
    not incompt(Data,Data1).

not cancelled1(P,Data) :-
    forall(P1,forall(Data1,not 
cancelled1(P,Data,P1,Data1))).

not cancelled(A,B) :-
    not cancelled1(A,B).

not stream1(A,_,X) :-
    A\=1.
not stream1(A,B,X) :-
    A=1,
    B\=p(X).

not stream1(A,B) :-
    forall(X,not stream1(A,B,X)).

not stream2(A,_) :-
    A\=2.
not stream2(A,B) :-
    A=2,
    B\=q(a).

not o_stream3(A,_) :-
	A\=2.
not o_stream3(A,B) :-
	A=2,
	B\=q(b).

not stream4(A,_) :-
    A\=3.
not stream4(A,B) :-
    A=3,
    B\=p(a).

not stream(A,B) :-
    not stream1(A,B),
    not stream2(A,B),
    not stream3(A,B),
    not stream4(A,B).

not valid_stream1(P,Data) :-
    not stream(P,Data).
not valid_stream1(P,Data) :-
    stream(P,Data),
    cancelled(P,Data).

not valid_stream(A,B) :-
    not valid_stream1(A,B).

not false.

nmr_check.
\end{lstlisting}
\end{multicols}

\subsection{s(CASP) output of stream.pl}

The next figure shows the output for the query
\code{?-valid_stream(Pr,Data)} when it is made to the program \texttt{stream.pl}
(see~\ref{app:appendix-stream}.1). The output to a query consists of:
(i) a justification tree with the successful derivation (note that
variables could be free, ground, or constrained); (ii)
a model with the positive atoms defined by the program that support
the successful derivation; and (iii) the bindings of variables in the
query (in this example, the bindings of \code{Pr} and \code{Data}). The
constraint store active at each call is shown close to each variable.

\begin{lstlisting}[style=tree]
?- valid_stream(Pr, Data).

Answer 1	(in 18.907 ms):

valid_stream(1,p({A.\=.[a,b]})) :-
   stream(1,p({A.\=.[a,b]})),
   not cancelled(1,p({A.\=.[a,b]})) :-
      not o_cancelled1(1,p({A.\=.[a,b]})) :-
         forall(B,forall(C,not o_cancelled1(1,p({A.\=.[a,b]}),B,C))) :-
            forall(C,not o_cancelled1(1,p({A.\=.[a,b]}),{D.=<.1},C)) :-
               not o_cancelled1(1,p({A.\=.[a,b]}),{D.=<.1},C) :-
                  not higher_prio({D.=<.1},1) :-
                     not o_higher_prio1({D.=<.1},1) :-
                        {D.=<.1}.=<.1.
            forall(C,not o_cancelled1(1,p({A.\=.[a,b]}),{E.>.3},C)) :-
               not o_cancelled1(1,p({A.\=.[a,b]}),{E.>.3},F) :-
                  higher_prio({E.>.3},1) :-
                     {E.>.3}.>.1.
                  not stream({E.>.3},F) :-
                     not o_stream1({E.>.3},F) :-
                        forall(G,not o_stream1({E.>.3},F,G)) :-
                           not o_stream1({E.>.3},F,G) :-
                              {E.>.3}\=1.
                     not o_stream2({E.>.3},F) :-
                        {E.>.3}\=2.
                     not o_stream3({E.>.3},F) :-
                        {E.>.3}\=2.
                     not o_stream4({E.>.3},F) :-
                        {E.>.3}\=3.
            forall(C,not o_cancelled1(1,p({A.\=.[a,b]}),{H.>.2, H.<.3},C)) :-
               not o_cancelled1(1,p({A.\=.[a,b]}),{H.>.2, H.<.3},I) :-
                  higher_prio({H.>.2, H.<.3},1) :-
                     {H.>.2, H.<.3}.>.1.
                  not stream({H.>.2, H.<.3},I) :-
                     not o_stream1({H.>.2, H.<.3},I) :-
                        forall(J,not o_stream1({H.>.2, H.<.3},I,J)) :-
                           not o_stream1({H.>.2, H.<.3},I,J) :-
                              {H.>.2, H.<.3}\=1.
                     not o_stream2({H.>.2, H.<.3},I) :-
                        {H.>.2, H.<.3}\=2.
                     not o_stream3({H.>.2, H.<.3},I) :-
                        {H.>.2, H.<.3}\=2.
                     not o_stream4({H.>.2, H.<.3},I) :-
                        {H.>.2, H.<.3}\=3.
            forall(C,not o_cancelled1(1,p({A.\=.[a,b]}),{K.>.1, K.<.2},C)) :-
               not o_cancelled1(1,p({A.\=.[a,b]}),{K.>.1, K.<.2},L) :-
                  higher_prio({K.>.1, K.<.2},1) :-
                     {K.>.1, K.<.2}.>.1.
                  not stream({K.>.1, K.<.2},L) :-
                     not o_stream1({K.>.1, K.<.2},L) :-
                        forall(M,not o_stream1({K.>.1, K.<.2},L,M)) :-
                           not o_stream1({K.>.1, K.<.2},L,M) :-
                              {K.>.1, K.<.2}\=1.
                     not o_stream2({K.>.1, K.<.2},L) :-
                        {K.>.1, K.<.2}\=2.
                     not o_stream3({K.>.1, K.<.2},L) :-
                        {K.>.1, K.<.2}\=2.
                     not o_stream4({K.>.1, K.<.2},L) :-
                        {K.>.1, K.<.2}\=3.
            forall(C,not o_cancelled1(1,p({A.\=.[a,b]}),2,C)) :-
               not o_cancelled1(1,p({A.\=.[a,b]}),2,{N.\=.[q(a),q(b)]}) :-
                  higher_prio(2,1) :-
                     2.>.1.
                  not stream(2,{N.\=.[q(a),q(b)]}) :-
                     not o_stream1(2,{N.\=.[q(a),q(b)]}) :-
                        forall(O,not o_stream1(2,{N.\=.[q(a),q(b)]},O)) :-
                           not o_stream1(2,{N.\=.[q(a),q(b)]},O) :-
                              2\=1.
                     not o_stream2(2,{N.\=.[q(a),q(b)]}) :-
                        2=2,
                       {N.\=.[q(a),q(b)]}\=q(a).
                     not o_stream3(2,{N.\=.[q(a),q(b)]}) :-
                        2=2,
                       {N.\=.[q(a),q(b)]}\=q(b).
                     not o_stream4(2,{N.\=.[q(a),q(b)]}) :-
                        2\=3.
               not o_cancelled1(1,p({A.\=.[a,b]}),2,q(a)) :-
                  proved(higher_prio(2,1)),
                  stream(2,q(a)),
                  not incompt(p({A.\=.[a,b]}),q(a)) :-
                     not o_incompt1(p({A.\=.[a,b]}),q(a)) :-
                        forall(P,not o_incompt1(p({A.\=.[a,b]}),q(a),P)) :-
                           not o_incompt1(p({A.\=.[a,b]}),q(a),{A.\=.[a,b]}) :-
                              p({A.\=.[a,b]})=p({A.\=.[a,b]}),
                              q(a)\=q({A.\=.[a,b]}).
                           not o_incompt1(p({A.\=.[a,b]}),q(a),a) :-
                              p({A.\=.[a,b]})\=p(a).
                     not o_incompt2(p({A.\=.[a,b]}),q(a)) :-
                        forall(P,not o_incompt2(p({A.\=.[a,b]}),q(a),P)) :-
                           not o_incompt2(p({A.\=.[a,b]}),q(a),P) :-
                              p({A.\=.[a,b]})\=q(P).
               not o_cancelled1(1,p({A.\=.[a,b]}),2,q(b)) :-
                  proved(higher_prio(2,1)),
                  stream(2,q(b)),
                  not incompt(p({A.\=.[a,b]}),q(b)) :-
                     not o_incompt1(p({A.\=.[a,b]}),q(b)) :-
                        forall(Q,not o_incompt1(p({A.\=.[a,b]}),q(b),Q)) :-
                           not o_incompt1(p({A.\=.[a,b]}),q(b),{A.\=.[a,b]}) :-
                              p({A.\=.[a,b]})=p({A.\=.[a,b]}),
                              q(b)\=q({A.\=.[a,b]}).
                           not o_incompt1(p({A.\=.[a,b]}),q(b),a) :-
                              p({A.\=.[a,b]})\=p(a).
                           not o_incompt1(p({A.\=.[a,b]}),q(b),b) :-
                              p({A.\=.[a,b]})\=p(b).
                     not o_incompt2(p({A.\=.[a,b]}),q(b)) :-
                        forall(R,not o_incompt2(p({A.\=.[a,b]}),q(b),R)) :-
                           not o_incompt2(p({A.\=.[a,b]}),q(b),R) :-
                              p({A.\=.[a,b]})\=q(R).
            forall(C,not o_cancelled1(1,p({A.\=.[a,b]}),3,C)) :-
               not o_cancelled1(1,p({A.\=.[a,b]}),3,{S.\=.[p(a)]}) :-
                  higher_prio(3,1) :-
                     3.>.1.
                  not stream(3,{S.\=.[p(a)]}) :-
                     not o_stream1(3,{S.\=.[p(a)]}) :-
                        forall(T,not o_stream1(3,{S.\=.[p(a)]},T)) :-
                           not o_stream1(3,{S.\=.[p(a)]},T) :-
                              3\=1.
                     not o_stream2(3,{S.\=.[p(a)]}) :-
                        3\=2.
                     not o_stream3(3,{S.\=.[p(a)]}) :-
                        3\=2.
                     not o_stream4(3,{S.\=.[p(a)]}) :-
                        3=3,
                       {S.\=.[p(a)]}\=p(a).
               not o_cancelled1(1,p({A.\=.[a,b]}),3,p(a)) :-
                  proved(higher_prio(3,1)),
                  stream(3,p(a)),
                  not incompt(p({A.\=.[a,b]}),p(a)) :-
                     not o_incompt1(p({A.\=.[a,b]}),p(a)) :-
                        forall(U,not o_incompt1(p({A.\=.[a,b]}),p(a),U)) :-
                           not o_incompt1(p({A.\=.[a,b]}),p(a),{A.\=.[a,b]}) :-
                              p({A.\=.[a,b]})=p({A.\=.[a,b]}),
                              p(a)\=q({A.\=.[a,b]}).
                           not o_incompt1(p({A.\=.[a,b]}),p(a),a) :-
                              p({A.\=.[a,b]})\=p(a).
                           not o_incompt1(p({A.\=.[a,b]}),p(a),b) :-
                              p({A.\=.[a,b]})\=p(b).
                     not o_incompt2(p({A.\=.[a,b]}),p(a)) :-
                        forall(V,not o_incompt2(p({A.\=.[a,b]}),p(a),V)) :-
                           not o_incompt2(p({A.\=.[a,b]}),p(a),V) :-
                              p({A.\=.[a,b]})\=q(V).
add_to_query :- o_nmr_check.

[ valid_stream(1,p({A.\=.[a,b]})), stream(1,p({A.\=.[a,b]})), higher_prio({E.>.3},1), 
  higher_prio({H.>.2, H.<.3},1), higher_prio({K.>.1, K.<.2},1), higher_prio(2,1), 
  stream(2,q(a)), stream(2,q(b)), higher_prio(3,1), stream(3,p(a)), o_nmr_check ]

Pr = 1,
Data = p({A.\=.[a,b]}) ? ;

Answer 2	(in 49.191 ms):

valid_stream(2,q(b)) :-
   stream(2,q(b)),
   not cancelled(2,q(b)) :-
      not o_cancelled1(2,q(b)) :-
         forall(B,forall(C,not o_cancelled1(2,q(b),B,C))) :-
            forall(C,not o_cancelled1(2,q(b),{A.=<.2},C)) :-
               not o_cancelled1(2,q(b),{A.=<.2},D) :-
                  not higher_prio({A.=<.2},2) :-
                     not o_higher_prio1({A.=<.2},2) :-
                        {A.=<.2}.=<.2.
            forall(C,not o_cancelled1(2,q(b),{E.>.3},C)) :-
               not o_cancelled1(2,q(b),{E.>.3},F) :-
                  higher_prio({E.>.3},2) :-
                     {E.>.3}.>.2.
                  not stream({E.>.3},F) :-
                     not o_stream1({E.>.3},F) :-
                        forall(G,not o_stream1({E.>.3},F,G)) :-
                           not o_stream1({E.>.3},F,G) :-
                              {E.>.3}\=1.
                     not o_stream2({E.>.3},F) :-
                        {E.>.3}\=2.
                     not o_stream3({E.>.3},F) :-
                        {E.>.3}\=2.
                     not o_stream4({E.>.3},F) :-
                        {E.>.3}\=3.
            forall(C,not o_cancelled1(2,q(b),{H.>.2, H.<.3},C)) :-
               not o_cancelled1(2,q(b),{H.>.2, H.<.3},I) :-
                  higher_prio({H.>.2, H.<.3},2) :-
                     {H.>.2, H.<.3}.>.2.
                  not stream({H.>.2, H.<.3},I) :-
                     not o_stream1({H.>.2, H.<.3},I) :-
                        forall(J,not o_stream1({H.>.2, H.<.3},I,J)) :-
                           not o_stream1({H.>.2, H.<.3},I,J) :-
                              {H.>.2, H.<.3}\=1.
                     not o_stream2({H.>.2, H.<.3},I) :-
                        {H.>.2, H.<.3}\=2.
                     not o_stream3({H.>.2, H.<.3},I) :-
                        {H.>.2, H.<.3}\=2.
                     not o_stream4({H.>.2, H.<.3},I) :-
                        {H.>.2, H.<.3}\=3.
            forall(C,not o_cancelled1(2,q(b),3,C)) :-
               not o_cancelled1(2,q(b),3,{K.\=.[p(a)]}) :-
                  higher_prio(3,2) :-
                     3.>.2.
                  not stream(3,{K.\=.[p(a)]}) :-
                     not o_stream1(3,{K.\=.[p(a)]}) :-
                        forall(L,not o_stream1(3,{K.\=.[p(a)]},L)) :-
                           not o_stream1(3,{K.\=.[p(a)]},L) :-
                              3\=1.
                     not o_stream2(3,{K.\=.[p(a)]}) :-
                        3\=2.
                     not o_stream3(3,{K.\=.[p(a)]}) :-
                        3\=2.
                     not o_stream4(3,{K.\=.[p(a)]}) :-
                        3=3,
                       {K.\=.[p(a)]}\=p(a).
               not o_cancelled1(2,q(b),3,p(a)) :-
                  proved(higher_prio(3,2)),
                  stream(3,p(a)),
                  not incompt(q(b),p(a)) :-
                     not o_incompt1(q(b),p(a)) :-
                        forall(M,not o_incompt1(q(b),p(a),M)) :-
                           not o_incompt1(q(b),p(a),M) :-
                              q(b)\=p(M).
                     not o_incompt2(q(b),p(a)) :-
                        forall(N,not o_incompt2(q(b),p(a),N)) :-
                           not o_incompt2(q(b),p(a),{O.\=.[b]}) :-
                              q(b)\=q({O.\=.[b]}).
                           not o_incompt2(q(b),p(a),b) :-
                              q(b)=q(b),
                              p(a)\=p(b).
add_to_query :- o_nmr_check.


[ valid_stream(2,q(b)), stream(2,q(b)), higher_prio({E.>.3},2), higher_
  prio({H.>.2, H.<.3},2), higher_prio(3,2), stream(3,p(a)), o_nmr_check ]

Pr = 2,
Data = q(b) ? ;

Answer 3	(in 1.606 ms):

valid_stream(3,p(a)) :-
   stream(3,p(a)),
   not cancelled(3,p(a)) :-
      not o_cancelled1(3,p(a)) :-
         forall(B,forall(C,not o_cancelled1(3,p(a),B,C))) :-
            forall(C,not o_cancelled1(3,p(a),{A.=<.3},C)) :-
               not o_cancelled1(3,p(a),{A.=<.3},D) :-
                  not higher_prio({A.=<.3},3) :-
                     not o_higher_prio1({A.=<.3},3) :-
                        {A.=<.3}.=<.3.
            forall(C,not o_cancelled1(3,p(a),{E.>.3},C)) :-
               not o_cancelled1(3,p(a),{E.>.3},F) :-
                  higher_prio({E.>.3},3) :-
                     {E.>.3}.>.3.
                  not stream({E.>.3},F) :-
                     not o_stream1({E.>.3},F) :-
                        forall(G,not o_stream1({E.>.3},F,G)) :-
                           not o_stream1({E.>.3},F,G) :-
                              {E.>.3}\=1.
                     not o_stream2({E.>.3},F) :-
                        {E.>.3}\=2.
                     not o_stream3({E.>.3},F) :-
                        {E.>.3}\=2.
                     not o_stream4({E.>.3},F) :-
                        {E.>.3}\=3.
add_to_query :- o_nmr_check.

[ valid_stream(3,p(a)), stream(3,p(a)), higher_prio({E.>.3},3), o_nmr_check ]

Pr = 3,
Data = p(a) ? ;

no
\end{lstlisting}

\section{Yale Scenario Example}
\label{app:yale-example}

\subsection{ASP + constraint encoding of yale\_shooting\_asp.pl}

Nest figure shows the spoiled Yale shooting scenario model written in
clingo + constraints using multi-shot
solving~\cite{janhunen2017clingo}.

\begin{multicols}{2}
\begin{lstlisting}[style=MyASP]
#include "incmode_lc.lp".
#program base.
action(load).
action(shoot).
action(wait).
duration(load,25).
duration(shoot,5).
duration(wait,36).
unloaded(0).
&sum { at(0) } = 0.
&sum { armed(0) } = 0.

#program step(n).
1 { do(X,n) : action(X) } 1.
&sum { at(n),-1*at(N') } = D :-
     do(X,n),
     duration(X,D),
     N' = n - 1.

loaded(n) :-
     loaded(n-1),
     not unloaded(n).
unloaded(n) :-
     unloaded(n-1),
     not loaded(n).
dead(n) :-
     dead(n-1).

&sum { armed(n) } = 0 :-
     unloaded(n-1).
&sum { armed(n),-1*armed(N') } = D :-
     do(X,n),
     duration(X,D),
     N' = n - 1,
     loaded(N').

loaded(n) :- 
     do(load,n).
unloaded(n) :- 
     do(shoot,n).
dead(n) :- 
     do(shoot,n), 
     &sum { armed(n) } <= 35.

:- do(shoot,n), unloaded(n-1).

#program check(n).
:- not dead(n), query(n).
:- not &sum {at(n)} <= 100, query(n).
:- do(shoot,n), not &sum {at(n)} > 35.    
\end{lstlisting}
\end{multicols}

\section{The Traveling Salesman Problem Example}
\label{app:trav-salesm-probl}

\subsection{ASP encoding of hamicycle\_asp.pl}

The next figure shows an ASP program for the Travelling Salesman
Problem described in section~\ref{sec:trav-salesman}. The encoding for
the Hamiltonian cycle part is from~\cite{dovier2005comparison} and the
code of \texttt{ \#sum} is adapted to run using \emph{clingo}.  The
bound on the total distance is one of the global constraints in the
program.

\begin{lstlisting}[style=MyASP]
1 {cycle(X,Y) : edge(X,Y)} 1 :- node(X).
1 {cycle(Z,X) : edge(Z,X)} 1 :- node(X).

reachable(X) :- node(X), cycle(b,X).
reachable(Y) :- node(X), node(Y), reachable(X), cycle(X,Y).

:- not reachable(X), node(X).

cycle_length(N) :- N = #sum {C: cycle(X,Y), distance(X, Y, C)}.
:- cycle_length(N), N >= 10.      % Cycles whose length is less than 10

edge(X,Y) :- distance(X,Y,D).
cycle_dist(U,V,D) :- cycle(U,V), distance(U,V,D).

node(a).         node(b).          node(c).            node(d).
distance(b,c,3).    %% ASP does not support rationals.
distance(c,d,8).    %% ASP does not support intervals.
distance(d,a,1).          distance(c,a,1).         distance(d,b,1).
distance(a,b,1).          distance(a,d,1).
\end{lstlisting}

\subsection{CLP(FD) encoding of hamicycle\_clpfd.pl}
\label{subapp:trav-salesman-fd}

The next figure shows the program in CLP($\mathit{FD}$) for the
Hamiltonian cycle problem presented in~\cite{dovier2005comparison},
using SICStus Prolog 3.11.2. Note that the library predicate
\mbox{\code{circuit/1}} does the bulk of the work.  Its implementation
is non-trivial and  shares a lot of code with the
implementation of \emph{all\_different}, and it implicitly imposes that
constraint. It does not calculate cycle lengths, but even in this
(simplified) case, the code as a whole is much larger that either the
ASP or s(CASP) code.

\begin{multicols}{2}
\begin{lstlisting}[style=MyProlog, basewidth=0.52em]
hamiltonian_path(Path) :- 
   graph(Nodes,Edges), 
   length(Nodes,N),
   length(Path,N), 
   domain(Path,1,N),
   make_domains(Path,1,Edges,N),
   circuit(Path), 
   labeling([ff],Path).

make_domains([],_,_,_).
make_domains([X|Y],Node,Edges,N) :-
   findall(Z,
     member([Node,Z],Edges),Succs),
   reduce_domains(N,Succs,X),
   Node1 is Node + 1, 
   make_domains(Y,Node1,Edges,N).

reduce_domains(0,_,_) :- !.
reduce_domains(N,Succs,Var) :- 
   N > 0, 
   member(N,Succs), !,
   N1 is N - 1, 
   reduce_domains(N1,Succs,Var).
reduce_domains(N,Succs,Var) :-
   Var #\= N, 
   N1 is N - 1,  
   reduce_domains(N1,Succs,Var).
\end{lstlisting}
\end{multicols}

\subsection{s(CASP) output of hamicycle\_scasp.pl}

The next figure shows the output to the query
\code{?-D.<.10,cycle(a,D,Cycle)} to the program hamicycle\_scasp.pl
(Figure~\ref{fig:travel-salesman} in Section~\ref{sec:trav-salesman}).

\begin{lstlisting}[style=tree]
?- D.<.10, travel_path(b, D, Cycle).

Answer 1	(in [2346.489] ms):

[ travel_path(b,61/10,[b,[31/10],c,[1],a,[1],d,[1],b]), path(b,b,b,61/10,[],
  [b,[31/10],c,[1],a,[1],d,[1],b]), cycle_dist(d,b,1), cycle(d,b), edge(d,b), 
  distance(d,b,1), node(d), node(b), node(a), edge(d,a), distance(d,a,1), 
  other(d,a), node(b), cycle(d,b), node(c), distance(d,b,1), path(b,b,d,51/10,
  [[1],b],[b,[31/10],c,[1],a,[1],d,[1],b]), cycle_dist(a,d,1), cycle(a,d), 
  edge(a,d), distance(a,d,1), edge(a,b), distance(a,b,1), other(a,b), node(d), 
  cycle(a,d), distance(a,d,1), path(b,b,a,41/10,[[1],d,[1],b],[b,[31/10],c,[1],
  a,[1],d,[1],b]), cycle_dist(c,a,1), cycle(c,a), edge(c,a), distance(c,a,1), 
  edge(c,d), distance(c,d,{A.>.8, A.<.21rat2}), other(c,d), node(a), cycle(c,a), 
  distance(c,a,1), path(b,b,c,31/10,[[1],a,[1],d,[1],b],[b,[31/10],c,[1],a,[1],
  d,[1],b]), cycle_dist(b,c,31/10), cycle(b,c), edge(b,c), distance(b,c,3.1), 
  distance(b,c,31/10), o_nmr_check, reachable(a), cycle(c,a), edge(c,a), 
  distance(c,a,1), reachable(b), cycle(d,b), edge(d,b), distance(d,b,1), 
  reachable(d), cycle(a,d), edge(a,d), distance(a,d,1), reachable(c), cycle(b,c), 
  edge(b,c), distance(b,c,3.1), other(a,a), node(d), other(a,c), node(d), 
  other(b,a), node(c), other(b,b), node(c), other(b,d), node(c), other(c,b), 
  node(a), other(c,c), node(a), other(d,c), node(b), other(d,d), node(b) ]

Cycle = [b,[31/10],c,[1],a,[1],d,[1],b],
D = 61/10 ? 
\end{lstlisting}

\section{Towers of Hanoi Example}
\label{app:towers-hanoi}

\subsection{ASP encoding of toh\_asp.pl}

The next program is part of~\cite{gebser2008user}:

\begin{lstlisting}[style=MyASP, basewidth=0.48em]
% Instance
peg(a;b;c).
disk(1..7).
init_on(1..7,a).
goal_on(1..7,b).
moves(127).
% Generate
1 { move(D,P,T) : disk(D) : peg(P) } 1 :- moves(M), T = 1..M.
% Define
move(D,T) :- move(D,_,T).
on(D,P,0) :- init_on(D,P).
on(D,P,T) :- move(D,P,T).
on(D,P,T+1) :- on(D,P,T), not move(D,T+1), not moves(T).
blocked(D-1,P,T+1) :- on(D,P,T), disk(D), not moves(T).
blocked(D-1,P,T) :- blocked(D,P,T), disk(D).
% Test
:- move(D,P,T), blocked(D-1,P,T).
:- move(D,T), on(D,P,T-1), blocked(D,P,T).
:- goal_on(D,P), not on(D,P,M), moves(M).
:- not 1 { on(D,P,T) : peg(P) } 1, disk(D), moves(M), T = 1..M.
#hide.
#show move/3.
\end{lstlisting}
\redaftlst

\subsection{ASP incremental encoding of toh\_aspI.pl}

The next program is part of the \emph{clingo} distribution and is
available at
\url{https://github.com/potassco/clingo/tree/master/examples/gringo/toh}

\begin{multicols}{2}
\begin{lstlisting}[style=MyASP, basewidth=0.48em]
#include <incmode>.

#program base.
peg(a;b;c).
disk(1..7).
init_on(1..7,a).
goal_on(1..7,b).

on(D,P,0) :- init_on(D,P).

#program step(t).
1 {move(D,P,t): disk(D),peg(P)} 1.

move(D,t) :- move(D,P,t).
on(D,P,t) :- move(D,P,t).
on(D,P,t) :- on(D,P,t-1), 
             not move(D,t).
blocked(D-1,P,t) :- on(D,P,t-1).
blocked(D-1,P,t) :- blocked(D,P,t), 
                    disk(D).
:- move(D,P,t), blocked(D-1,P,t).
:- move(D,t), on(D,P,t-1), blocked(D,P,t).
:- not 1 { on(D,P,t) } 1, disk(D).

#program check(t).
:- query(t), goal_on(D,P), not on(D,P,t).

#show move/3.
\end{lstlisting}
\end{multicols}

\subsection{s(CASP) output of hanoi.pl}
 
\begin{lstlisting}[style=tree]
?- hanoi(7,T).

Answer 1	(in [420.343] ms):

[ hanoi(7,127), move(a,b,1), move(a,c,2), move(b,c,3), move(a,b,4), 
  move(c,a,5), move(c,b,6), move(a,b,7), move(a,c,8), move(b,c,9), 
  move(b,a,10), move(c,a,11), move(b,c,12), move(a,b,13), move(a,c,14), 
  move(b,c,15), move(a,b,16), move(c,a,17), move(c,b,18), move(a,b,19), 
  move(c,a,20), move(b,c,21), move(b,a,22), move(c,a,23), move(c,b,24), 
  move(a,b,25), move(a,c,26), move(b,c,27), move(a,b,28), move(c,a,29), 
  move(c,b,30), move(a,b,31), move(a,c,32), move(b,c,33), move(b,a,34), 
  move(c,a,35), move(b,c,36), move(a,b,37), move(a,c,38), move(b,c,39), 
  move(b,a,40), move(c,a,41), move(c,b,42), move(a,b,43), move(c,a,44), 
  move(b,c,45), move(b,a,46), move(c,a,47), move(b,c,48), move(a,b,49), 
  move(a,c,50), move(b,c,51), move(a,b,52), move(c,a,53), move(c,b,54), 
  move(a,b,55), move(a,c,56), move(b,c,57), move(b,a,58), move(c,a,59), 
  move(b,c,60), move(a,b,61), move(a,c,62), move(b,c,63), move(a,b,64), 
  move(c,a,65), move(c,b,66), move(a,b,67), move(c,a,68), move(b,c,69), 
  move(b,a,70), move(c,a,71), move(c,b,72), move(a,b,73), move(a,c,74), 
  move(b,c,75), move(a,b,76), move(c,a,77), move(c,b,78), move(a,b,79), 
  move(c,a,80), move(b,c,81), move(b,a,82), move(c,a,83), move(b,c,84), 
  move(a,b,85), move(a,c,86), move(b,c,87), move(b,a,88), move(c,a,89), 
  move(c,b,90), move(a,b,91), move(c,a,92), move(b,c,93), move(b,a,94), 
  move(c,a,95), move(c,b,96), move(a,b,97), move(a,c,98), move(b,c,99), 
  move(a,b,100), move(c,a,101), move(c,b,102), move(a,b,103), move(a,c,104), 
  move(b,c,105), move(b,a,106), move(c,a,107), move(b,c,108), move(a,b,109), 
  move(a,c,110), move(b,c,111), move(a,b,112), move(c,a,113), move(c,b,114), 
  move(a,b,115), move(c,a,116), move(b,c,117), move(b,a,118), move(c,a,119), 
  move(c,b,120), move(a,b,121), move(a,c,122), move(b,c,123), move(a,b,124), 
  move(c,a,125), move(c,b,126), move(a,b,127) ]

T = 127 ?
\end{lstlisting}

\end{document}